\documentclass[twocolumn]{aastex631}

\makeatletter
\DeclareRobustCommand{\HI}{%
  \mbox{H\check@mathfonts\fontsize\sf@size\z@\selectfont I}%
}
\makeatother

\usepackage{amsmath}

\begin{document}


\title{A Break In the Size-Stellar Mass Relation: \\ Evidence for Quenching and Feedback in Dwarf Galaxies}

\author[0000-0002-1598-5995]{Nushkia Chamba}
\thanks{NASA Postdoctoral Program Fellow}
\affiliation{NASA Ames Research Center,  Space Science and Astrobiology Division M.S. 245-6, Moffett Field, CA 94035, USA}
\affiliation{The Oskar Klein Centre, Department of Astronomy, Stockholm University, AlbaNova, SE-10691 Stockholm, Sweden}
\correspondingauthor{Nushkia Chamba}
\email{nushkia.chamba@nasa.gov}

\author{Pamela M. Marcum}
\affiliation{NASA Ames Research Center,  Space Science and Astrobiology Division M.S. 245-6, Moffett Field, CA 94035, USA}

\author{Am\'elie Saintonge}
\affiliation{Department of Physics \& Astronomy, University College London, London, UK, WC1E 6BT}


\author[0000-0003-3249-4431]{Alejandro S. Borlaff} 
\affiliation{NASA Ames Research Center, Space Science and Astrobiology Division M.S. 245-6, Moffett Field, CA 94035, USA}
\affiliation{Bay Area Environmental Research Institute, Moffett Field, California 94035, USA}

\author{Matthew J. Hayes}
\affiliation{The Oskar Klein Centre, Department of Astronomy, Stockholm University, AlbaNova, SE-10691 Stockholm, Sweden}

\author[0000-0002-5714-799X]{Valentin J. M. Le Gouellec}
\thanks{NASA Postdoctoral Program Fellow}
\affiliation{NASA Ames Research Center,  Space Science and Astrobiology Division M.S. 245-6, Moffett Field, CA 94035, USA}


\author[0000-0003-0346-6722]{S. Drew Chojnowski}
\thanks{NASA Postdoctoral Program Fellow}
\affiliation{NASA Ames Research Center,  Space Science and Astrobiology Division M.S. 245-6, Moffett Field, CA 94035, USA}

\author{Michael N. Fanelli}
\affiliation{NASA Ames Research Center,  Space Science and Astrobiology Division M.S. 245-6, Moffett Field, CA 94035, USA}






\begin{abstract}

Mapping stars and gas in nearby galaxies is fundamental for understanding their growth and the impact of their environment. This issue is addressed by comparing the stellar `edges’ of galaxies $D_{\rm stellar}$, defined as the outermost diameter where in situ star formation significantly drops, with the gaseous distribution parameterized by the neutral atomic hydrogen diameter measured at 1\,$M_{\odot}$/pc$^2$, $D_{\HI{}}$.  By sampling a broad \HI{} mass range $10^5\,M_{\odot} < M_{\HI{}} < 10^{11}\,M_{\odot}$, we find several dwarf galaxies with $M_{\HI{}} < 10^9\,M_{\odot}$ from the field and Fornax Cluster which are distinguished by $D_{\rm stellar} >> D_{\HI{}}$. For the cluster dwarfs, the average \HI{} surface density near $D_{\rm stellar}$ is $\sim$0.3\,$M_{\odot}$/pc$^2$, reflecting the impact of quenching and outside-in gas removal from ram pressure and tidal interactions. In comparison, $D_{\rm stellar}/D_{\HI{}}$ ranges between 0.5--2 in dwarf field galaxies, consistent with the expectations from stellar feedback.  Only more massive disk galaxies in the field can thus be characterized by the common assumption that $D_{\rm stellar} \lesssim D_{\HI{}}$. We discover a break in the $D_{\rm stellar}-M_{\star}$ relation at $m_{break} \sim 4\times10^8\,M_{\odot}$ that potentially differentiates the low mass regime where the influence of stellar feedback and environmental processes more prominently regulates the sizes of nearby galaxies. Our results highlight the importance of combining deep optical and \HI{} imaging for understanding galaxy evolution.

\end{abstract}



\keywords{Scaling relations (2031) -- Galaxy radii (617) -- Galaxy environments (2029) -- Circumgalactic medium (1879)}


\section{Introduction} \label{sec:intro}


A truncation is a characteristic feature in the outer regions of spiral galaxies first discovered by \citet{vanderkruit1979aap38_15}. Observationally, this feature is  identified with a sharp cut-off or change in slope in the radial stellar light profiles of galaxies. One explanation for the origin of truncations is the existence of a gas density threshold for star formation \citep[typically defined using \HI{} surface density, see][]{kennicutt1989apj344_685, schaye2004apj609_667, leroy+2008aj136_2782}. The idea that the stellar boundary of a galaxy is limited by such a threshold has been proposed as a physically motivated galaxy size definition \citep{trujillo+2020mnras493_87}. This more general concept of stellar `edges', i.e. `the outermost radial location where in situ star formation significantly drops' \citep{chamba+2022aap667_87}, is supported by the radial cut-offs observed in the ultra-violet profiles of disk galaxies which is a sensitive tracer of ongoing star formation \citep[e.g.][]{martinezlombilla+2019mnras483_664}. However, \citet{chamba+2022aap667_87} has shown that similar features can be found using the optical colour profiles, allowing the stellar edge concept to be broadly applicable for a wider range of galaxy types than those originally studied by van~der~Kruit, including ellipticals and dwarfs.  \par 

Interestingly, when used as a size definition, \citet{chamba+2023arXiv2311.10144} have recently shown that the edge radius $R_{\rm edge}$ versus stellar mass $M_{\star}$ plane has a very small intrinsic scatter ($\sim~$0.07~dex) for galaxies with $10^5\,M_{\odot} < M_{\star} < 10^{11}M_{\odot}$. This value is typically two or three times smaller than the scatter from conventional size measures such as the effective radius $r_e$ \citep{trujillo+2020mnras493_87}. \citet{sanchezalmeida2020mnras495_78} show that the small scatter in the size-stellar mass plane can be reproduced when size is measured as a fixed surface density or brightness level \citep[see also the work by][]{hall+2012mnras425_2741, stone+2021apj912_41}. The small scatter is shown to be a consequence of the profile shapes of galaxies when approximated using the \citet{sersic1968book} law. There are three main reasons as to why the arguments made in \citet{sanchezalmeida2020mnras495_78} are not directly applicable to edge scaling relations.\par 

First, the edge radius does not occur at a fixed density or surface brightness \citep{chamba+2022aap667_87}. In the field, dwarf galaxies have edges where on average $\Sigma_{\star}(R_{
\rm edge})\sim 0.5\,M_{\odot}$/pc$^2$, spirals $\sim 1\,M_{\odot}$/pc$^2$ and massive ellipticals $\sim 3\,M_{\odot}$/pc$^2$. Second, when size is measured as a fixed surface density or brightness level, the scatter in the size--stellar mass plane significantly increases in the dwarf regime $M_{\star} < 10^{9}\,M_{\odot}$ as the central regions of dwarfs become more diffuse and the measure tends towards zero \citep[see][]{watkins+2023mnras521_2012}. And third, \citet{sanchezalmeida2020mnras495_78} use single \citet{sersic1968book} models in their analysis, while real galaxies are far more complex. Therefore, the small scatter found in \citet{chamba+2023arXiv2311.10144} must be additionally explained by other physical mechanisms. \par 

In fact, \citet{chamba+2023arXiv2311.10144} show that the scatter in the $R_{\rm edge}-M_{\star}$ plane is driven by the morphology and environment of galaxies. At a fixed stellar mass in the range $10^5\,M_{\odot} < M_{\star} < 10^{10}\,M_{\odot}$, Fornax cluster galaxies can be up to $50\%$ smaller and denser at $R_{\rm edge}$ compared to nearly isolated field galaxies. However, the environmental difference is stronger for early-type dwarfs compared to the late-types. Previous work using classical optical parameters such as $r_e$ which encloses half the total light of a galaxy \citep{devaucouleurs1948afz11_247} often miss the outer regions of galaxies, and have thus been unable to trace the impact of environment on the comprehensive distribution of stellar emission associated with these galaxies \citep[e.g.][]{derijcke+2009mnras393_798, huertascompany+2013apj779_29}.   
\par 
Measuring the sizes of galaxies accurately is also important for understanding the distribution of halo masses in dark matter models \citep[e.g.][]{wechsler+2018araa56_435} as well as the growth of galaxies over cosmic time \citep[e.g.][]{buitrago+2024aap682_110}. These open questions have traditionally been addressed using optical size parameters, mainly due to the wide availability of large, multi-wavelength surveys and galaxy samples from the ground \citep[e.g. the Sloan Digital Sky Survey, or SDSS;][]{york+2000aj120_1579} and space \citep[e.g. HST/CANDELS;][]{koekemoer+2011apj197_36}. However, the advent of deep neutral atomic hydrogen (\HI{}) observations has also allowed detailed analysis of the \HI{} size--mass plane for large samples of galaxies \citep[e.g.][]{wang+2016mnras460_2143, rajohnson+2022mnras512_2697}. 

\HI{} is a key component of the interstellar medium and environment needed for star formation in galaxies \citep{saintonge+2022araa60_319}. Typically, the \HI{} diameter $D_{\HI{}}$ is defined as the location of the 1\,$M_{\odot}$/pc$^2$ surface density isomass contour, inspired by the optical `$D_{25}$' \citep{bosma1981aj86_1825}. $D_{25}$ is an optical diameter definition for galaxies at the location of the 25\,mag arcsec$^{-2}$ isophote in the $B$-band, dating back to \citet{redman1936mnras96_588} \citep[but see the discussion on \emph{`\HI{} extent'} and a \emph{`vain attempt to find a good characteristic radius'} in][]{bosma1981aj86_1825}. \par 

Remarkably, the $D_{\HI{}}-M_{\HI{}}$ relation \citep{broeils+1997aap324_877} has been shown to hold for galaxies in all environments, including clusters \citep{wang+2016mnras460_2143}. \citet[][]{stevens+2019mnras490_96} show that this observation can be explained because when galaxies lose $\HI{}$ gas as a consequence of ram pressure or other environmental processes, the \HI{} redistributes in such a way that they still follow the $D_{\HI}-M_{\HI{}}$ relation. Consequently, the scatter of this relation remains small ($\lesssim$ 0.06\,dex), even at higher redshifts \citep{rajohnson+2022mnras512_2697}. Interestingly, this value is comparable to the optical $R_{\rm edge}-M_{\star}$ relation.  However, it is currently unknown if and how these relations are physically connected, given that the scatter in the stellar size-mass scaling relation is more sensitive to the environment of galaxies \citep{chamba+2023arXiv2311.10144}.  
\par

\begin{figure*}[th!]
    \centering
    \includegraphics[width=0.9\textwidth]{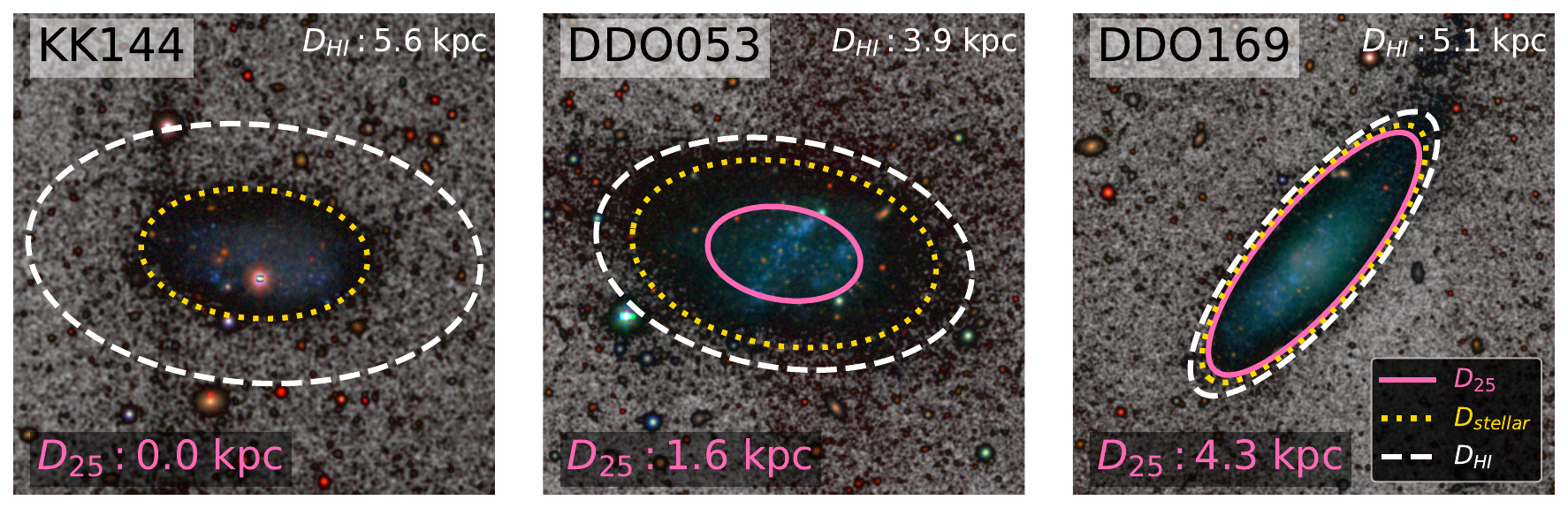}
    \caption{Variation of diameter $D_{25}$ in nearby dwarf galaxies with similar stellar mass $M_{\star} \sim 10^7\,M_{\odot}$ (or $M_{\HI{}} \sim 10^8\,M_{\odot}$). RGB images are created using DECaLs \citep{dey+2019aj157_168} which are overlaid on a grey scale to highlight the outer regions of the galaxy. Darker regions are faint. $D_{25}$ traces radically different stellar regions of galaxies (pink ellipse) compared to $D_{stellar}$, the latter showing the more consistent size measure expected for galaxies of $\sim$equal stellar masses. The misleading measure of the stellar disk extent complicates the interpretation of scaling relations or parameters based on $D_{25}$. Other diameter measures relevant to this study, the \HI{} diameter $D_{\HI{}}$ (white) and stellar edge $D_{\rm stellar}$ (yellow), are shown for comparison. $D_{25}$ and $D_{\HI{}}$ are taken from \citet{wang+2016mnras460_2143} while $D_{\rm stellar}$ from \citet{chamba+2023arXiv2311.10144}. The symbol `D' for all size measurements refer to diameters (i.e. twice the radius).}
    \label{fig:D25_examples}
\end{figure*}

Previous studies exploring the connection between the \HI{} and stellar properties of galaxies have used
size measures based on both light concentration similar to $r_e$ and the $D_{25}$ isophote \citep[e.g.][and references therein]{denes+2014mnras444_667, wang+2020apj890_63, jaiswal+2020mnras498_4745, reynolds+2023pasa40_032}. Unfortunately, $D_{25}$ also has several shortcomings. As $D_{25}$ is an isophote defined using the $B$-band, the measure is heavily dependent upon the location of recent star formation and dust in the galaxy. For example, star formation can occur in the center or within the disk, consequently depending on the galaxy's morphology and the shape of it's potential.  The challenges of using isophotes to represent a self-consistent size metric are not immediately remedied by simply using a redder filter that is less sensitive to these components.  An isophotal size based on any wavelength inherently traces light concentration, and therefore can fail to represent the extent of galaxies with diffuse light at large galactic radii.  \par 

For instance, $D_{25}$ may not always represent the outer stellar distribution of dwarf galaxies where the central surface brightness can be near 25\,mag arcsec$^{-2}$ \citep[see][]{simon2019araa57_375}. We visualise this problem for three nearby dwarf galaxies with $M_{\star} \sim 10^7\,M_{\odot}$ in Fig. \ref{fig:D25_examples}. In one case, $D_{25}$ is zero because it is at the center of the dwarf. \par 

This practical limitation is often ignored in the interpretations of scaling relations, even though it may lead to incorrect conclusions \citep[see][]{chamba+2020aap633_3}. Unlike previous measures, the edge is a common, physical feature in the outskirts of galaxies which is not impacted by light concentration or image depth.  This work explores the impact of using the stellar edges from \citet{chamba+2023arXiv2311.10144} to understand the $D_{\HI{}}-M_{\HI{}}$ relation for gas-rich nearly isolated and Fornax Cluster galaxies in their sample. \par 

\citet{chamba+2023arXiv2311.10144} additionally found evidence that galaxies with low \HI{} fractions have edges with high stellar surface density (see their Fig. 16). In the case of Fornax Cluster members, this result potentially indicates that the removal of \HI{} in the cluster environment due to tidal forces and ram pressure quenches galaxies outside-in \citep[e.g.][]{bekki2009mnras399_2221}, strongly impacting the location and density at which stellar edges are found. To investigate if this origin scenario is responsible for how galaxies populate the stellar and \HI{} size-mass relations, we compare the stellar and \HI{} diameters of galaxies quantified by the stellar edge (here notated as $D_{\rm stellar} = 2\times R_{\rm edge}$) and $D_{\HI{}}$ respectively. We specifically study the distribution of these size measures as a function of both $M_{\HI{}}$ and $M_{\star}$. The limitations of the currently used $D_{\HI{}}$ isophote will also be discussed. \par

The paper is organized as follows. Section \ref{sect:data} presents the data, catalogues, and measurements we obtained from the literature. Section \ref{sect:method} summarizes our methodology for radial profile derivation, edge and HI size measurement. Section \ref{sect:scaling} presents our main results using the HI and stellar edge--mass relations and we discuss them in Section \ref{sect:discussion}. We assume a flat $\Lambda$-CDM cosmology with $H_{0} = 70$\, km s$^{-1}$ Mpc$^{-1}$ and $\Omega=0.3$.

\section{Data and Sample} \label{sect:data}

An overview of all the data and catalogues used in this work is presented in Table \ref{tab:data_source}. Sect. \ref{hi_sample} describes the main \HI{} sample selected for analysis from \citet{chamba+2023arXiv2311.10144} for comparing galaxies in the Fornax Cluster with those in less dense groups and isolated environments. The \citet{chamba+2023arXiv2311.10144} catalogue consists of the $D_{\rm stellar}$ and $M_{\star}$ measurements. A brief summary of how $D_{\rm stellar}$ and $M_{\star}$ are estimated is provided in Sect. \ref{sect:method}. The next Sect. \ref{useful_ref} is a compilation of useful reference samples from the literature which are used to define the $D_{\HI{}}-M_{\HI{}}$ size scaling relation and verify the robustness of the methods used in this work i.e. the scaling relations must be applicable to a wide range of galaxies and not be specific to measurement methodologies of edges or sample selection biases. 

\begin{table*}[tbh!]
    \begin{center}
    \caption{Data and sample overview}

    \begin{tabular}{cccc}
        \hline 
        \hline 
         & $M_{\HI{}}$ range & No. of &   \\ 
       Sample & ($M_{\odot}$)  & galaxies & References  \\ & & &  \\ \hline 
        
       \underline{\textit{Analysis samples}}: & & & \\ 
       (i) Fornax Cluster  & $10^5-10^{10}$ & 23 & \citet[ATCA;][]{loni+2021aap648_31}, \citet[MeerKAT;][]{kleiner+2023aap675_108} \\

       (ii) Field: Groups \& & $10^5 - 10^{11}$ & 91 & \citet{karachentsev+2013aj145_101}, \citet{durbala+2020aj160_271}, \\
       isolated galaxies & & & \& \citet{zhu+2023mnras521_3765}  \\ 
        & & & \\

       \underline{\textit{Reference samples}}: & & & \\ 
       (i) Gas-rich spirals \& dwarfs & $10^7 - 10^{11}$  & 407 & \citet{wang+2016mnras460_2143} \& references therein \\ 

       & & & \\
       
       (ii) Nearby disk galaxies with  & $10^8-10^{10}$ & 29  & \citet{vanderkruit2007aap466_883}, \citet{trujillo+2016apj823_123}, \\ 
        identified truncations &  &  & \citet{martinezlombilla+2019mnras483_664}, \citet{trujillo+2021aap654_40}, \\
        & & & \citet{diazgarcia+2022aap667_109} \& \citet{ossafuentes+2023apj951_149} \\ 

        & & & \\
        
       (iii) Extreme LSB galaxies & $10^8$, $10^{10}$& 2 & \citet{montes+2024aap681_15}, \citet{hagen+2016apj826_210} \\

       \hline 
       
    \end{tabular} 
    \end{center}
    \textbf{Notes.} The main \HI{} analysis samples are selected from \citet{chamba+2023arXiv2311.10144}. $D_{\rm stellar}$ is reported in \citet{chamba+2023arXiv2311.10144} and reference sample (ii). Reference sample (i) is mainly used to define the $D_{\HI{}}-M_{\HI{}}$ relation and sample (iii) have the radial profiles necessary to visually measure $D_{\rm stellar}$. The listed references for each sample indicate the source catalogue used to obtain \HI{} mass (and/or $D_{\HI{}}$ if available) measurements.
    \label{tab:data_source}
\end{table*}

\begin{figure*}[thb!]
    \centering
    \includegraphics[width=1.0\textwidth]{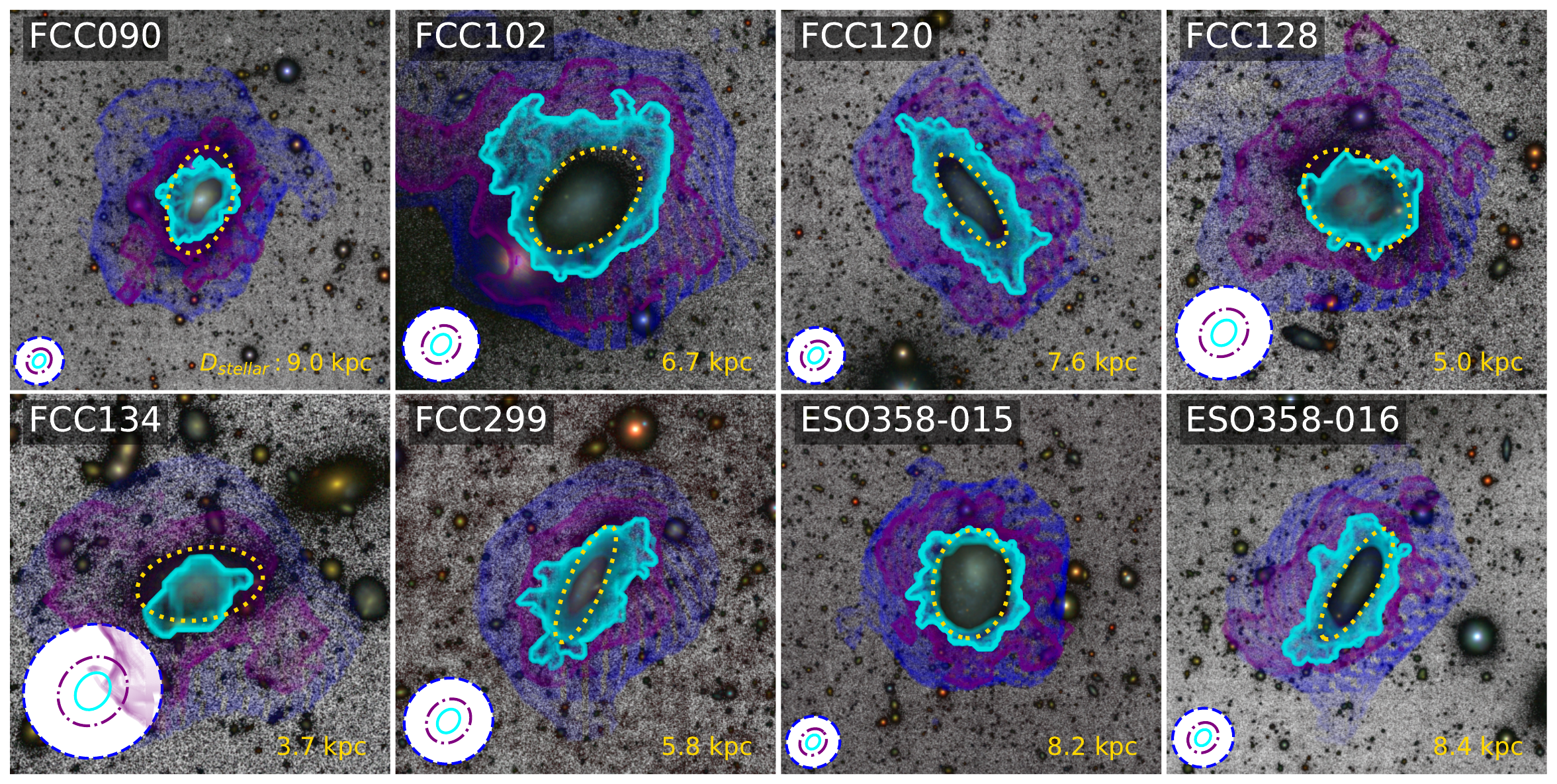}
    \caption{Visualisation of a sub-sample of \HI{} maps from the MeerKAT Fornax Survey. These maps are displayed by drawing equally spaced, two hundred isomass contours between the range $0.01-1\,M_{\odot}$/pc$^2$, overlaid on the FDS optical RGB images. All contours are of equal thickness. The three colours of the \HI{} data correspond to three levels of spatial resolution and \HI{} sensitivity which are publicly available. From highest to lowest resolution: cyan (labelled as 11\arcsec{} in the survey products which is near its beam size), purple (21\arcsec{}) and blue (41\arcsec{}) have column density limits ($3\sigma$ over 25\,km/s) of 5$\times10^{19}$, 1.2$\times10^{19}$ and 2.7$\times10^{18}$ cm$^{-2}$, respectively \citep[Table 2 in][]{serra+2023aap673_146}.  These beam sizes are plotted as ellipses on the left corner of each panel. The RGB images  were created in the same way as in Fig. \ref{fig:D25_examples}. The innermost contour of the \HI{} map in cyan corresponds to the contour at 1\,$M_{\odot}$/pc$^2$ and the outermost in blue is the 0.01\,$M_{\odot}$/pc$^2$ level. \HI{} surface densities \emph{below} the 1\,$M_{\odot}$/pc$^2$ isocontour are chosen as this threshold is used to define $D_{\HI{}}$. This choice highlights the significant extent of \HI{} at  lower column densities and to provide context to $D_{\HI}$. We draw the $D_{\rm stellar}$ measure using a dotted yellow ellipse as a marker of the stellar boundary. The inner 1\,$M_{\odot}$/pc$^2$ contour is not always significantly outside this stellar boundary in some of the galaxies shown.}
    \label{fig:mfs}
\end{figure*}

\subsection{\HI{} detections in the Fornax Cluster and field}
\label{hi_sample}
\textbf{\HI{} in the Fornax Cluster:} The Fornax Cluster is a nearby (20\,Mpc), low mass \citep[$M_{virial} \sim 5\times10^{13}\,M_{\odot}$;][]{drinkwater+2001mnras326_1076} cluster. About 850 galaxies have been catalogued in Fornax, including 275 low surface brightness dwarfs \citep{su+2021aap647_100, venhola+2022aap662_43}. Given its proximity, the cluster has been observed in multiple wavelengths. The 21-cm \HI{} line is no exception. \par 
The $M_{\HI{}}$ of the gas-rich Fornax cluster galaxies are taken from two surveys. The first is the blind \HI{} survey using the Australia Telescope Compact Array \citep[ATCA;][]{loni+2021aap648_31} which targets the center of the cluster out to its virial radius. The integrated \HI{} maps are created using a combination of over 300\,hrs of observations divided into 756 pointings. The spatial resolution of the survey is $95 \times 67${\arcsec} and has a $3\sigma$ \HI{} column density sensitivity of $2\times 10^{19}$cm$^{-2}$ or $M_{\HI{}} \sim 2\times 10^7\,M_{\odot}$ at the distance of Fornax. \HI{} was detected in sixteen out of the 200 spectroscopically confirmed Fornax cluster members. We refer the reader to \citet{loni+2021aap648_31} for more details. \par 
The second is the ongoing MeerKAT Fornax Survey \citep[MFS;][]{serra+2023aap673_146}. The survey targets the center of the cluster and in-falling south-west sub-group (Fornax A) over 91 pointings. While a slightly smaller area coverage than the ATCA survey, MeerKAT provides \HI{} maps with a much higher spatial resolution $12.2 \times 9.6$ arcsec and depth, reaching a $3\sigma$ $M_{\HI{}}$ sensitivity of $\sim 6\times10^5\,M_{\odot}$ at the distance of Fornax. This sensitivity is ideal for the study of gas-rich dwarfs in the cluster. In addition to the \HI{} rich galaxies found by \citet{loni+2021aap648_31}, \citet{kleiner+2023aap675_108} reported new \HI{} detections in an additional 8 dwarf galaxies: five of late-type morphology and three of early-type. \par 
For this work, the 23 gas-rich Fornax galaxies which were previously studied in \citet{chamba+2023arXiv2311.10144} are selected. All except one galaxy in this sub-sample are of late-type morphology (see the Fornax Deep Survey \citep[FDS;][]{peletier+2020arXiv2008.12633} catalogues \citet{venhola+2018aap620_165, su+2021aap647_100}). The \HI{} maps of the MeerKAT Fornax Survey\footnote{\protect\url{https://sites.google.com/inaf.it/meerkatfornaxsurvey/data}} are shown for a sub-sample of these galaxies in Fig. \ref{fig:mfs}. The low and high resolution maps provided by the MFS collaboration which are created using different beam sizes \citep[see Table 2 in][]{serra+2023aap673_146} are overlaid to visualise the full range of the data. The lower resolution maps capture the extended, low column density \HI{}. To highlight the detection of \HI{} at these  lower column densities, only regions which have \HI{} surface densities \emph{below} 1\,$M_{\odot}$/pc$^2$, i.e. the threshold used to defined $D_{\HI{}}$, are shown in the figure panels. In other words, the innermost contour of the highest spatial resolution cyan \HI{} map represents the contour at 1\,$M_{\odot}$/pc$^2$.\par  


\textbf{\HI{} in group and isolated environments:} For comparison with the Fornax Cluster environment, \HI{} detected galaxies in 
nearly isolated and group environments (the field) are selected from \citet{chamba+2022aap667_87, chamba+2023arXiv2311.10144}, a large sample of nearby ($z < 0.1$) galaxies for which $D_{\rm{stellar}}$ and $M_{\star}$ measurements are available. The sample consists of elliptical, spiral and dwarf galaxies with $10^5\,M_{\odot} < M_{\star} < 10^{12}\,M_{\odot}$ in the footprints of the SDSS Stripe 82 \citep{york+2000aj120_1579, fliri+2016mnras456_1359}, Dark Energy Camera Legacy Surveys \citep[DECaLs;][]{dey+2019aj157_168} and FDS \citep[][]{peletier+2020arXiv2008.12633}. \par 

Specifically, the group satellite galaxies are taken from the SAGA \citep{geha+2017apj847_4} and ELVES \citep{carlsten+2021apj922_267} surveys and nearly isolated galaxies from \citet{karachentsev+2013aj145_101} and \citet{chamba+2022aap667_87}. The satellite and isolated galaxies are jointly called field galaxies. \par 
Total \HI{} masses are available only for a sub-sample of field galaxies in this sample. They are compiled using \citet{karachentsev+2013aj145_101} who estimate the \HI{} masses using integrated $\HI{}$ fluxes collected from various literature sources, the majority from the \HI{} Parkes All Sky Survey \citep{meyer+2004mnras350_1195} and Arecibo \citep[ALFALFA;][]{giovanelli+2005aj130_2598}, the ALFALFA+SDSS catalog \cite{durbala+2020aj160_271} and \citet{zhu+2023mnras521_3765} who study group satellite galaxies using ALFALFA. We find 91 \HI{} detected galaxies from \citet{chamba+2023arXiv2311.10144} with $10^5\,M_{\odot} < M_{\HI{}} < 10^{11}\,M_{\odot}$. All except one are of late-type morphology. \par 

\subsection{Reference samples from the literature}
\label{useful_ref}

\textbf{I. Gas-rich spirals and dwarfs:} To define the $D_{\HI{}}-M_{\HI{}}$ relation, we use the \citet{wang+2016mnras460_2143} sample as the reference for this work. This sample is a compilation of 418 galaxies using 15 different \HI{} surveys, observed using the Very Large Array, ATCA and Westerbork Synthesis Radio Telescope to name a few.  The survey targets are predominantly gas-rich spiral and dwarf irregular galaxies with \HI{} masses between $10^7\,M_{\odot} < M_{\HI{}} < 10^{11}\,M_{\odot}$ and environments from field to clusters. The cluster member galaxies are from Ursa Major \citep{verheijen+2001aap370_765} and Virgo \citep{chung+2009aj138_1741} surveys.  \par

The $D_{\HI{}}$,  $D_{25}$ and $M_{\HI{}}$ of this sample are directly taken from the \cite{wang+2016mnras460_2143} catalogue. The stellar masses $M_{\star}$ of the sample  are not reported in \citet{wang+2016mnras460_2143}. For this reason, we compile their $M_{\star}$ using a combination of several public catalogues (see Table \ref{tab:wang_stellar}) to verify that the sample spans a similar range in $M_{\star}$ as the \citet{chamba+2023arXiv2311.10144} one. This compilation provided $M_{\star}$ estimations for 407 out of the 418 unique galaxies in \cite{wang+2016mnras460_2143}.  These galaxies have stellar masses between $10^6\,M_{\odot} < M_{\star} < 10^{12}\,M_{\odot}$ which brackets the range of the \citet{chamba+2022aap667_87, chamba+2023arXiv2311.10144} sample. \par\par 

\begin{table*}[htb!]
    \begin{center}
    \caption{Stellar mass compilation}
    \begin{tabular}{p{10cm}p{2.2cm}}
    \hline 
    \hline
       Source catalogue  &  No. of matches \\ \hline
       \begin{enumerate}
       \item $z=0$ Multi-wavelength Galaxy Synthesis \citep{leroy+2019apj244_24} 
    \item Updated Nearby Galaxy Catalogue \citep{karachentsev+2013aj145_101} 
       \item GALEX-SDSS-WISE Legacy Catalogue \citep[A2\footnote{See \protect\url{https://salims.pages.iu.edu/gswlc/}};][]{salim+2016apj227_2} 
       \item ALFALFA-SDSS \citep{durbala+2020aj160_271} 
       \item LITTLE THINGS \citep{zhang+2012aj143_47}, LVHIS \citep{wang+2017mnras472_3029} \& WHISP \citep{naluminsa+2021mnras502_5711} 
       \end{enumerate} & 

       \begin{enumerate} 
       \item[] 237 
       \item[] 58 
       \item[] 58 
       \item[] 21 
       \item[] 23
       \end{enumerate} \\
    \hline 
    
    \end{tabular}
    
    \end{center}
    \textbf{Notes.} Compiling the stellar masses of galaxies in the \citet[][]{wang+2016mnras460_2143} catalogue. Galaxies which were not in catalogues 1-4 were then matched to those listed in 5. The total sample falls in the stellar mass range $10^6\,M_{\odot} < M_{\star} < 10^{12}\,M_{\odot}$, bracketing the \citet{chamba+2022aap667_87, chamba+2023arXiv2311.10144} sample.
    \label{tab:wang_stellar}
\end{table*}





\textbf{II. Truncations reported in nearby disks:} To complement the above samples in the higher mass regime, we additionally retrieve the truncations of disk galaxies studied by \citet{vanderkruit2007aap466_883}, \citet{martinezlombilla+2019mnras483_664}, \citet{trujillo+2016apj823_123, trujillo+2021aap654_40}, \citet{diazgarcia+2022aap667_109} and \citet{ossafuentes+2023apj951_149}. All except the galaxies studied in \citet{trujillo+2016apj823_123, trujillo+2021aap654_40} are edge-on in configuration and have warped \HI{} distributions. \par 
These authors use different datasets and methods than those used here to identify the truncation feature. Additionally, the galaxies themselves have not been previously examined in the context of size-mass scaling relations.  Therefore, these individual data points can serve as a test for the robustness of the methodology and comparison explored here. \par 
The authors of \citet{vanderkruit2007aap466_883}, \citet{martinezlombilla+2019mnras483_664} and \citet{diazgarcia+2022aap667_109} did not report the stellar and \HI{} masses of the galaxies they study. For this reason, we retrieved the masses of those galaxies from the aforementioned catalogues or the literature. The list of galaxies and references for their stellar masses are: NGC~4244 \citep{huchtmeier1973aap23_93}, NGC~4013 \citep{bottema1995aap295_605}, NGC~5907 \citep{dumke+1997aap325_124}, NGC~891 \citep{oosterloo+2007aj134_1019} and NGC~4565 \citep{heald+2011aap526_118}.\par

\textbf{III. Extremely low surface brightness galaxies}: To illustrate extreme galaxies in the size-mass relations, we also include the giant low surface brightness disk galaxy UGC~1382  \citep{hagen+2016apj826_210} and a very extended diffuse dwarf called `Nube' \citep{montes+2024aap681_15} in our sample. These authors did not report a truncation feature, but they did provide the radial profiles necessary to estimate its location visually, see \citet[Fig. 4;][]{hagen+2016apj826_210} and \citet[Fig. 5;][]{montes+2024aap681_15}. Future deep surveys from the LSST-Rubin and Roman may detect similar galaxies in larger numbers which will allow a more in-depth analysis of the size scaling relations of extremely faint galaxies compared to that of ``normal'' galaxies shown here. \par

\section{Methodology}
\label{sect:method}

This section summarizes how diameters are measured in the literature. Relevant to this work is the measurement of $D_{\HI{}}$ using the MeerKAT Fornax Survey and the identification of $D_{\rm stellar}$ from \citet{chamba+2023arXiv2311.10144}. For the $D_{\HI{}}$ of Fornax Cluster galaxies, we follow a similar procedure as in \citet{wang+2016mnras460_2143}. All other measurements when available, including the $D_{\HI{}}$, $M_{\HI{}}$, $D_{\rm stellar}$ and $M_{\star}$ are taken from the literature (Table \ref{tab:data_source} and \ref{tab:wang_stellar}). \par 

\subsection{Measuring $D_{\HI{}}$}
The highest resolved MeerKAT data (the $11''$ maps) are used to derive the radial \HI{} surface density profiles by first converting flux to surface densities using Eq. (50) from \citet{meyer+2017pasa34_52} and then azimuthally averaging the pixels over ellipitical bins of increasing distance beginning from the centre of the galaxy. The center and ellipses are defined using the same parameters as in \citet{chamba+2023arXiv2311.10144} which was used to measure $D_{\rm stellar}$. The mean density values in each ellipical bin are calculated using $3\sigma$-clipping. The major axis of the ellipse where the \HI{} surface density profile reaches 1\,$M_{\odot}/$pc$^2$ surface density is then determined as $D_{\HI{}}$. We apply the Gaussian approximation correction to account for beam smearing i.e. $D_{\HI{}, cor} = \sqrt{D_{\HI{}}^2 - Bmaj\times Bmin}$, $Bmaj$ and $Bmin$ are the major and minor axes of the beam \citep[see][]{wang+2016mnras460_2143}. Only galaxies where $D_{\HI{}}^2/ (Bmaj\times Bmin) > 1$ are selected to ensure that $D_{\HI{}}$ is resolved. The average value of this ratio considering the higher resolution datasets in MeerKAT \HI{}  maps 11\arcsec{} and 21\arcsec{} is 4.1.\par 
While not provided in \citet{wang+2016mnras460_2143}, the typical uncertainty in $D_{\HI{}}$ for nearby galaxies $z < 0.1$ is $\pm 0.02-0.11$\,kpc \citep[see][]{rajohnson+2022mnras512_2697}. These values were estimated using half the beam major axis of their data and the uncertainty in their inclination measurements. In the case of the MeerKAT survey used here, at the distance of the Fornax Cluster, the uncertainty in measuring $D_{\HI{}}$ using the 11\arcsec resolution maps is $\sim \pm 0.6$\,kpc. 

\subsection{Identifying $D_{\rm stellar}$}
For the identification of edges ($D_{\rm stellar}$), we refer the reader to the detailed examples and explanations provided by \citet{chamba+2022aap667_87, chamba+2023arXiv2311.10144}. One of the main criteria used for the identification of the edge location is the outermost inflection point or change in slope in the $g-r$ colour profile. This metric is useful especially when the feature is not clearly marked by a change in slope in the radial stellar light profile.  However, the individual surface brightness and stellar density profiles are also used to identify the outermost change in slope as this edge feature can appear in any one of these profiles. \par 
The full identification procedure is thus iterative, including the visualization of each 1D profile and the ellipse used to derive the radial profile at the initially selected location of the edge on the 2D RGB image of the galaxy. This step ensures that any unmasked or asymmetric features in the outer part of the galaxy (such as streams) are not incorrectly reported as the edge feature. Examples of asymmetric galaxies are discussed in detail in  \citet[Fig. 1 in][]{chamba+2022aap667_87} and \citet[Appendix B in][]{chamba+2023arXiv2311.10144}. It is shown that 1) the use of an ellipse to derive radial profiles is sensitive to the edge feature even in cases where there are streams and 2) the edge features reported are not a consequence of masking these features.  We use the above procedure to identify the edges in reference sample (iii) using the profiles published in \citet{hagen+2016apj826_210} and \citet{montes+2024aap681_15}. All other $D_{
\rm stellar}$ values are taken from \citet{chamba+2022aap667_87, chamba+2023arXiv2311.10144}. \par 

Regarding the quality of this visualization procedure and the subjective identification of edges, the typical scatter of repeated and iterative measurements is $\sim 0.04$\,dex. This value is the dispersion of the separate identification of edges (characterized by the selected $R_{\rm edge}$ value) by different authors in \cite{chamba+2022aap667_87} using the procedure outlined above. \citet{chamba+2023arXiv2311.10144} performed a similar analysis for the full sample of $\sim 1000$ galaxies and arrived at a similar dispersion value. Appendix B of \citet{chamba+2022aap667_87} additionally details the effect of image depth on these measurements and shows that all the edge locations reported are within the depth of the data used. Although it is beyond the scope of this work, other authors have recently shown that this vizualization procedure can also be automatized using image deep learning techniques \citep{fernandeziglesias+2024aap683_145}. \par  
In the case of $D_{\rm stellar}$, the main sources of uncertainty impacting the size--stellar mass relation are the image background and the galaxy stellar mass estimations. $M_{\star}$ is estimated by converting the $g-r$ color profile to a mass-to-light ratio following the relations calibrated by \citet{roediger+2015mnras452_3209} for the $g$-band using the \citet{bruzual+2003mnras344_1000} stellar population synthesis models (see their Appendix A, Table 2). The stellar surface density profile is then integrated until the surface brightness limit of the data used is reached. These estimates are provided in \citet{chamba+2023arXiv2311.10144} (see their Sect. 4 and Table 3). For the sub-sample analyzed here, the average uncertainty in the size--stellar mass relation due to the background is $\pm 0.057$\,dex and stellar mass estimation is $\pm 0.066$\,dex. These values correspond to a total uncertainty of $\sim$0.5\,kpc at the distance of the Fornax Cluster.



\section{Scaling relations} \label{sect:scaling}

\noindent The size scaling relations of the above sample are described in the following subsections.

\subsection{Diameter $D_{\HI{}}$ vs. $M_{\HI{}}$}

\begin{figure}[h]
    \centering
    \includegraphics[width=0.5\textwidth]{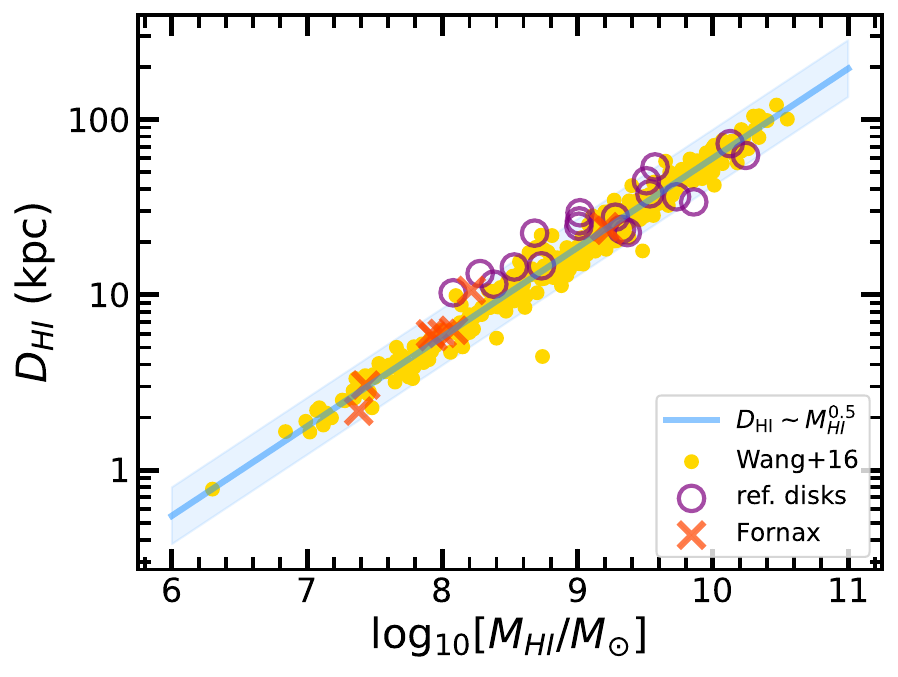}
    \caption{HI diameter-mass relation. The shaded region corresponds to $\sim 0.16$\,dex which is three times the 
    scatter 
    found by 
    \citet{rajohnson+2022mnras512_2697}.
    From the literature, measurements for gas-rich dwarfs and spirals from \citet{wang+2016mnras460_2143} (yellow points) and the \citet{vanderkruit2007aap466_883} reference disk sample (purple circles) are shown. The best-fit line is taken from \citet{wang+2016mnras460_2143}. MeerKAT \HI{} detected Fornax cluster galaxies analysed in this work are plotted as orange crosses.} 
    \label{fig:hi_mass_relation}
\end{figure}

Fig. \ref{fig:hi_mass_relation} shows the $D_{\HI{}}-M_{\HI{}}$ relation. The linear best fit is taken from \citet{wang+2016mnras460_2143}. The individual $D_{\HI{}}$ data points are taken from \citet{wang+2016mnras460_2143} (yellow dots) and \citet{vanderkruit2007aap466_883} for galaxies with reported truncations (purple rings). The $D_{\HI{}}$ for the Fornax cluster galaxies (this work) using MeerKAT data are plotted as crosses. While not reproduced here, \citet{loni+2021aap648_31} show that the ATCA Fornax sample also follows the $D_{\HI{}}-M_{\HI{}}$ relation within $3\sigma$ (see their Fig. 5). \par 
From the Fornax sample, there are two galaxies which do not appear on the relation. FCC128 is not resolved in our dataset according to our criteria (see Sect. \ref{sect:method}) and FCC134 is resolved but does not reach the $1\,M_{\odot}$/pc$^2$ density needed to measure $D_{\HI{}}$. Interestingly, FCC134 has a very low $M_{\HI} \sim 2\times10^6\,M_{\odot}$ \citep{kleiner+2023aap675_108} and the only gas-rich early-type dwarf in our sample. \par 





\subsection{$D_{\rm stellar}$ vs. $M_{\star}$}


\begin{figure}[t!]
    \centering
    \includegraphics[width=0.5\textwidth]{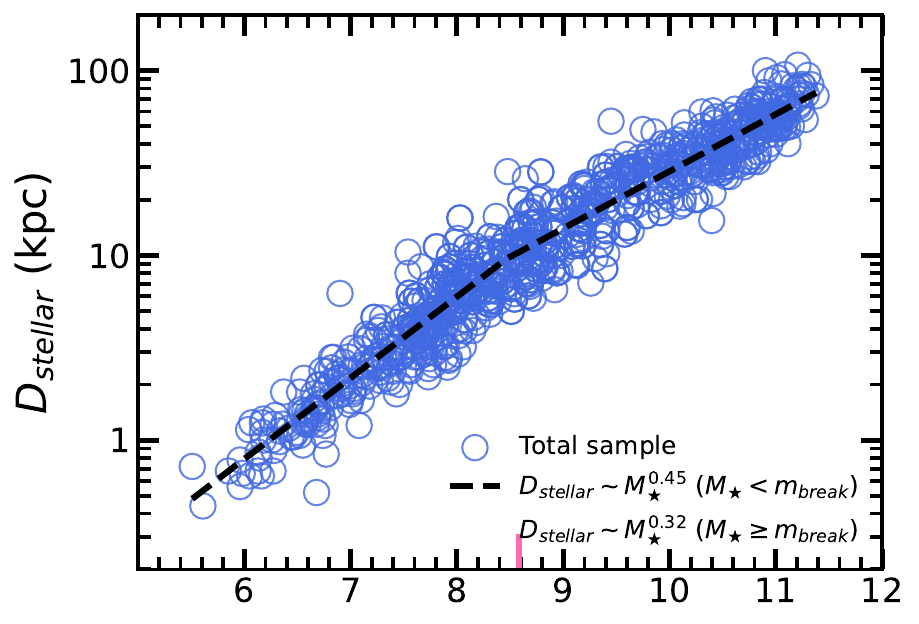}
    \includegraphics[width=0.51\textwidth]{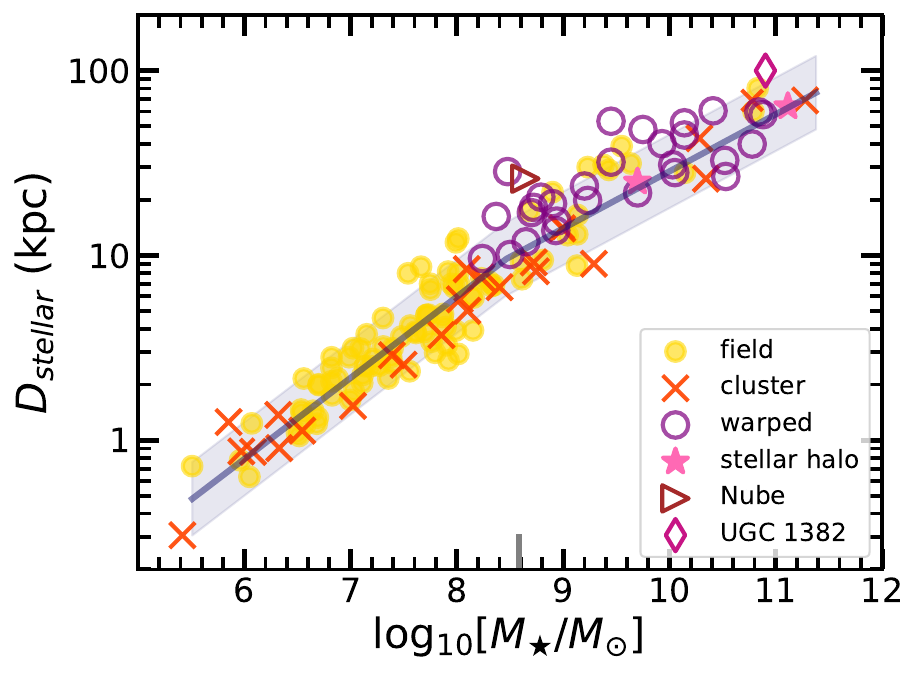}
    \caption{Edge diameter as a function of stellar mass. \emph{Upper:} Relation using the spirals and dwarfs sample from \citet{chamba+2022aap667_87, chamba+2023arXiv2311.10144} and reference disk sample (ii; see Table \ref{tab:data_source}). The best fit is piece wise linear, with a break at $m_{break} \sim 4\times 10^{8}\,M_{\odot}$. \emph{Lower:} Relation showing individual sub-samples for comparison with the Fornax Cluster galaxies, including the reference disk galaxy sample grouped here according to reported features: with warps and stellar halos (see Sect. \ref{useful_ref} for details). The two extreme low surface brightness galaxies \emph{Nube} \citep{montes+2024aap681_15} and \emph{UGC 1382} \citep{hagen+2016apj826_210} are shown separately. All except two galaxies in these sub-samples are of late-type morphology. The shaded region corresponds to $\sim 0.2$\,dex which is three times the intrinsic scatter of the relation.}
    \label{fig:dedge_mstar}
\end{figure}

The upper panel of Fig. \ref{fig:dedge_mstar} shows the $D_{\rm stellar}-M_{\star}$ plane for the spirals and dwarfs sample from \citet{chamba+2022aap667_87, chamba+2023arXiv2311.10144} and the reference sample from the literature (ii; see Table \ref{tab:data_source}). The lower panel shows the gas-rich galaxies selected for this work (Table \ref{tab:data_source}). Truncation measures for individual galaxies from the literature are labelled in the legend. The shaded region corresponds to three times the intrinsic scatter of the relation ($\sigma_{int} \sim 0.07$\,dex). Several remarks follow: \par 
\begin{itemize}

\item First, the best fit relation is piecewise linear, with a break at $m_{break} = 10^{8.6\pm 0.5}\,M_{\odot}$ and $D_{break} = 12.3\pm0.2$\,kpc: $D_{\rm stellar} \sim M^{0.45\pm0.03}_{\star}$ where $M_{\star} < m_{break}$ and $D_{\rm stellar} \sim M^{0.32\pm 0.02}_{\star}$ for galaxies with higher stellar mass. $m_{break}$ is obtained by using a piece wise linear fit routine in python\footnote{\protect\url{https://jekel.me/piecewise_linear_fit_py/}} \citep{jekel+2019manual} where the break point in the scaling relation is unknown and the optimum $m_{break}$ is such that the sum of least squares is minimized. The fit is constrained to solve for two line segments with a single breaking point. This break was verified using a bootstrap method for which a uniform random sub-sample of 300 galaxies from the parent sample was selected and fit 100 times. The quoted $m_{break}$ and slopes are the average of these repeated fits. The non-linear p-value of this piecewise linear model with a single break point is very small ($p \sim 10^{-170}$). Across the full stellar mass range, the observed dispersion is $\sigma_{obs}\sim 0.13$\,dex. Fits of the relation using other functional forms are shown in Appendix A.\par 

The best fit lines shown in Fig. \ref{fig:dedge_mstar} are:


\[
    \log(D_{\rm stellar}) = 
\begin{cases}
    0.45 \log(M_{\star}) - 2.78, & M_{\star} < m_{break} \\
    0.32 \log(M_{\star}) - 1.64,              & \text{otherwise}
\end{cases}
\]

The uncertainties in the intercepts are 0.18 and 0.16 respectively.

\item Second, not all gas-rich Fornax members (orange crosses) are below or above the $D_{\rm stellar}-M_{\star}$ relation. 
\item Third, even though the literature values obtained make use of different datasets and methods for identifying the truncation, all the data points shown are within 3$\sigma_{obs}$. Notably, the presence of warps or stellar halo-like features in these galaxies as well as the extremely low surface brightness galaxies are not major outliers in this relation. \par 

\end{itemize}



\subsection{$D_{25}$ and $D_{\rm stellar}$ vs. $M_{\HI{}}$}

\begin{figure}[h!]
    \centering
    \includegraphics[width=0.5\textwidth]{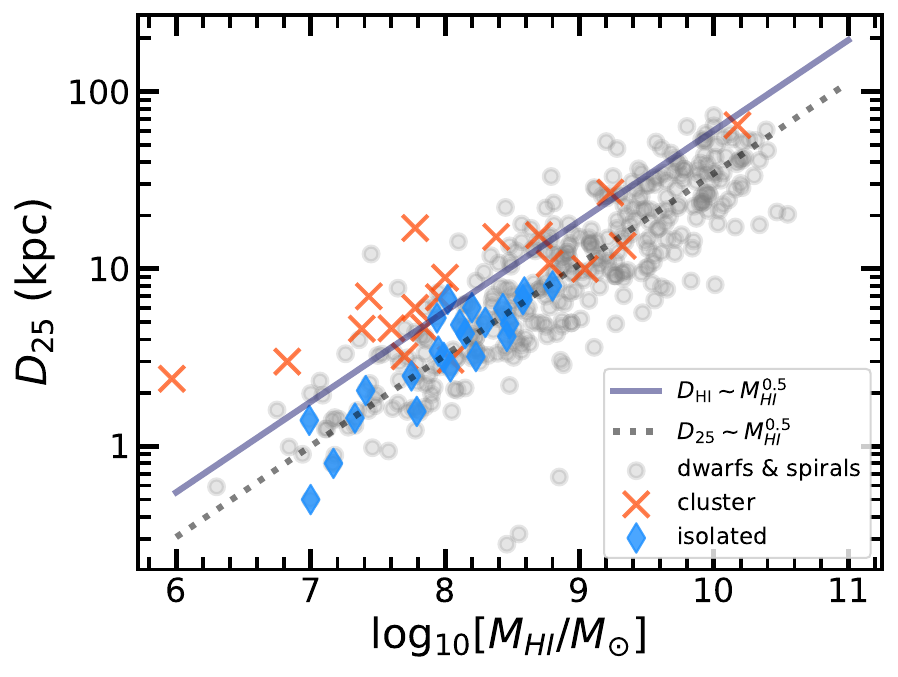}
    \caption{Diameter $D_{25}$ as a function of $M_{HI}$. Individual points in grey show the $D_{25}$ of the gas-rich dwarf and spiral \citet{wang+2016mnras460_2143} sample, orange crosses show the Fornax cluster galaxies from \citet{loni+2021aap648_31} and blue diamonds are a sub-sample of the isolated dwarfs from \citet{chamba+2023arXiv2311.10144} where $D_{25}$ measurements are available. The best fit lines for $D_{\HI{}}$ (solid, dark purple) and $D_{25}$ (dotted, grey) are taken from \citet{wang+2016mnras460_2143} and \citet{broeils+1997aap324_877}, respectively. The common statement in the literature that \HI{} is twice more extended than the optical distribution of galaxies is rooted in this comparison \citep[see][]{broeils+1997aap324_877}.}
    \label{fig:DHI_D25}
\end{figure}


In the classical work by \citet{broeils+1997aap324_877}, $D_{25}$ is compared to $D_{\HI{}}$ as shown in Fig. \ref{fig:DHI_D25}. This plot includes the fitted $D_{\HI{}}-M_{\HI{}}$ relation and $D_{25}$ for the \citet{wang+2016mnras460_2143} sample in the same plane. The $D_{25}-M_{\HI{}}$ best-fit from \citet{broeils+1997aap324_877} is also plotted and we find excellent agreement, with a scatter of $\sim 0.2\,$dex. The ratio between $D_{\HI{}}/D_{25}$ is $\sim1.7$, which has been the justification behind the statement in the literature  that \HI{} can extend two times more than the optical stellar distribution of galaxies. Available $D_{25}$ measurements for the gas-rich field and Fornax cluster sub-samples from \citet{chamba+2023arXiv2311.10144} in this plane are also labelled. \par 


\begin{figure*}[t]
    \centering
    \includegraphics[width=0.95\textwidth]{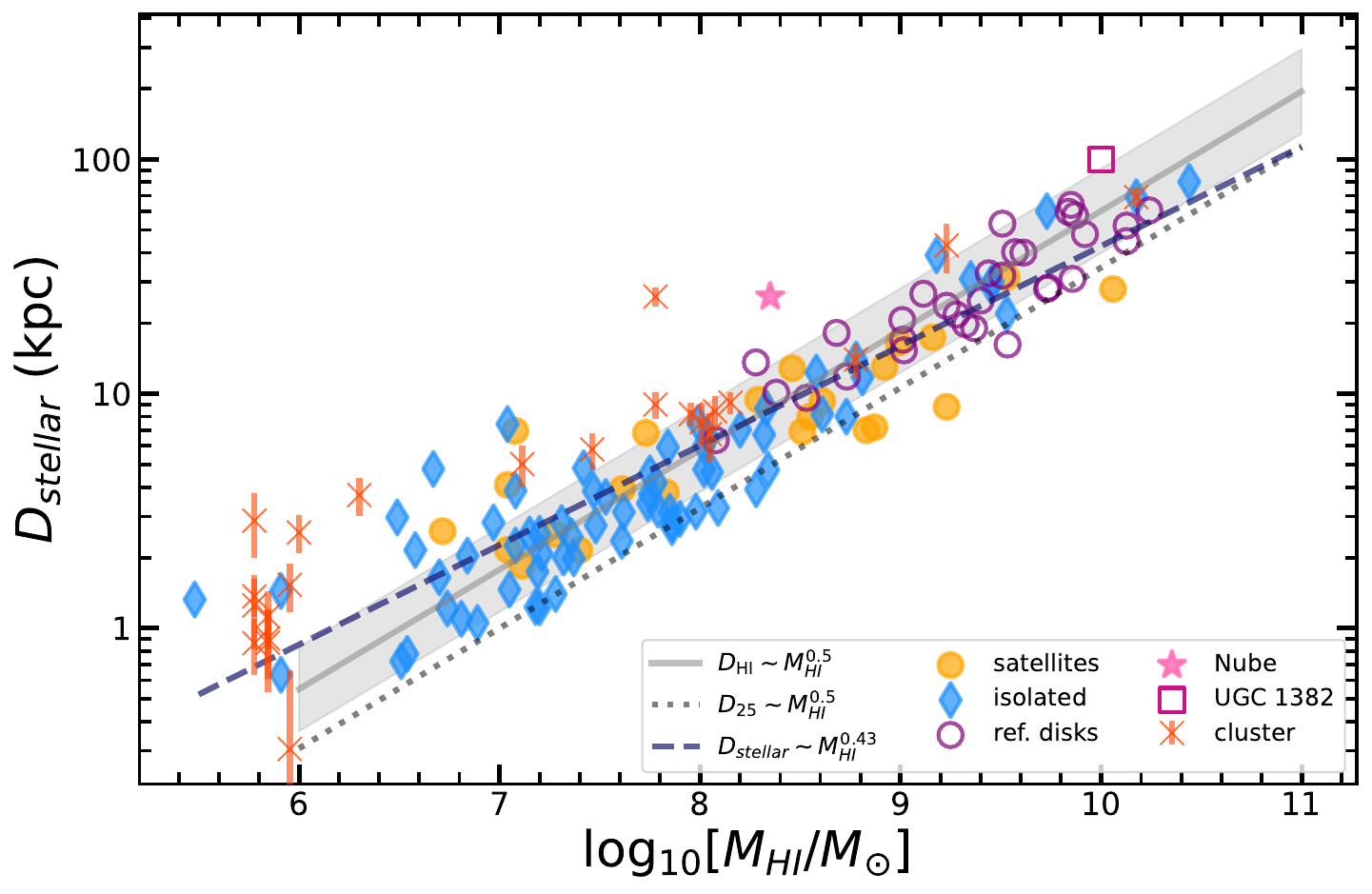}\\
    \caption{Stellar edge diameter $D_{\rm stellar}$ as a function of $M_{\HI{}}$ for the \HI{} detected galaxies in the sample. Compared to the isolated (blue diamonds) and satellite samples (yellow dots), all except one Fornax cluster galaxy (orange crosses)  are found above the best fit line $D_{\rm stellar} \sim M_{\HI{}}^{0.43}$ (dashed dark purple). The data points for the reference disk galaxy sample with warps or stellar halo features are collectively shown using purple rings. The $D_{\HI{}}$ (solid grey) and $D_{25}$ (dotted grey) lines from Fig. \ref{fig:DHI_D25} are plotted for comparison. The shaded region in the $D_{\HI{}}$ line shows the $3\sigma$ scatter to highlight galaxies with $D_{\rm stellar} >> D_{\HI{}}$ at a fixed $M_{\HI{}}$. Uncertainties are only shown for the cluster sample.}
    \label{fig:Dedge_MHI}
\end{figure*}

For comparison, Fig. \ref{fig:Dedge_MHI} shows the $D_{\rm stellar}$ vs. $M_{\HI{}}$ plane for the full gas-rich sample, including those for which $D_{25}$ measurements are unavailable. The $D_{\HI{}}$ and $D_{25}$ lines from Fig. \ref{fig:DHI_D25} are plotted for comparison. The different samples of galaxies considered in this work are labelled in the legend. For visualization reasons, uncertainties in the $D_{\rm stellar}$ measurement from \citet{chamba+2023arXiv2311.10144} are plotted only for the cluster sample. The typical total uncertainty of the measurements shown due to the background and stellar mass estimations are $\sim \pm 0.5$\,kpc \citep[see][for detailed computations]{chamba+2023arXiv2311.10144}.  Several remarks are in order:
\begin{itemize}
\item First, the diameter based on the edge scale as $D_{\rm stellar} \sim M_{\HI{}}^{0.43}$ with a scatter of $\sim 0.2\,$dex, a shallower slope than that of the $D_{\HI{}}-M_{\HI{}}$ relation. This result contrasts with \citet{broeils+1997aap324_877} $D_{25}-M_{\HI{}}$ fit (see Fig. \ref{fig:DHI_D25}). \par 

\item Second, for the isolated galaxies, typically $D_{\rm stellar}/D_{25} \sim 1.4\pm0.4$ and for the cluster sub-sample $D_{\rm stellar}/D_{25} \sim 1.2\pm0.1$. While $D_{25}$ is more similar to $D_{\rm stellar}$ in our cluster sub-sample, the  difference and variation in the measurements are significant for the isolated sample.
\item Third, although with a larger scatter, the slope of the relation $D_{\rm stellar} \sim M_{\HI{}}^{0.43}$ is similar to the size--stellar mass plane $D_{\rm stellar} \sim M_{\star}^{0.42}$ published in \citet{chamba+2023arXiv2311.10144} for galaxies for which $M_{\star} < 10^{10}\,M_{\odot}$. The observed scatter of $\sim 0.20$\,dex in the $D_{\rm stellar} \sim M_{\HI{}}^{0.43}$ relation is driven by less massive $M_{\HI{}} < 10^9\,M_{\odot}$ galaxies. The scatter is $\sim 0.14\,$dex for higher mass galaxies and similar to the observed $D_{\rm stellar} \sim M_{\star}^{0.42}$ relation. These values correspond to a difference of $\sim 39\%$.

\item Fourth, when fixing $M_{\HI{}}$, all except one of the Fornax cluster dwarf galaxies in our sample have their stellar edges beyond their $D_{\HI{}}$ (light grey solid line).These galaxies also have $g-r$ colours $\geq 0.5$. According to the $D_{\HI{}}-M_{\HI{}}$ relation, their $D_{\rm stellar} > D_{\HI{}}$ by a factor of 2.3$\pm1.3$. \par 
These findings are highlighted in Fig. \ref{fig:Dstellar_DHI_ratio} and Fig. \ref{fig:hi_profiles}. Fig. \ref{fig:Dstellar_DHI_ratio} shows the ratio $D_{\rm stellar}/D_{\HI{}}$ as a function of $M_{\HI{}}$ (left) and $M_{\star}$ (right) for galaxies in the sample where both size measurements are available. Fig. \ref{fig:hi_profiles}  shows the mean \HI{} profile of the MeerKAT Fornax cluster sample analyzed in this work using the 11\arcsec{} and 21\arcsec{} resolution maps. The \HI{} surface density near $D_{\rm stellar}$ is $\sim 0.3\,M_{\odot}/$pc$^2$, more than a factor of three lower compared to the $1\,M_{\odot}$/pc$^2$ level.


\begin{figure*}
    \centering
    \includegraphics[width=0.5\textwidth]{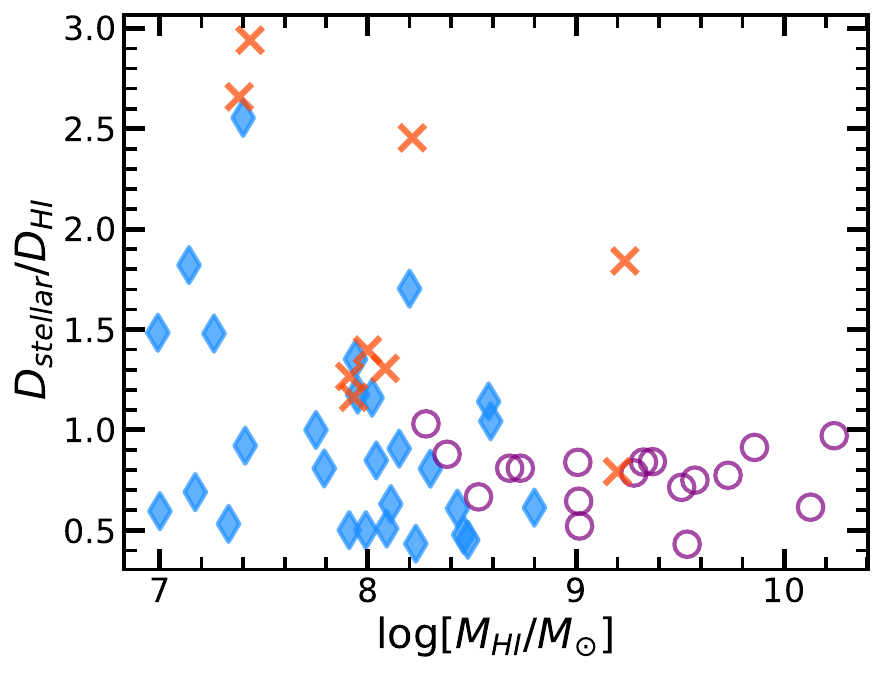} 
    \includegraphics[width=0.472\textwidth]{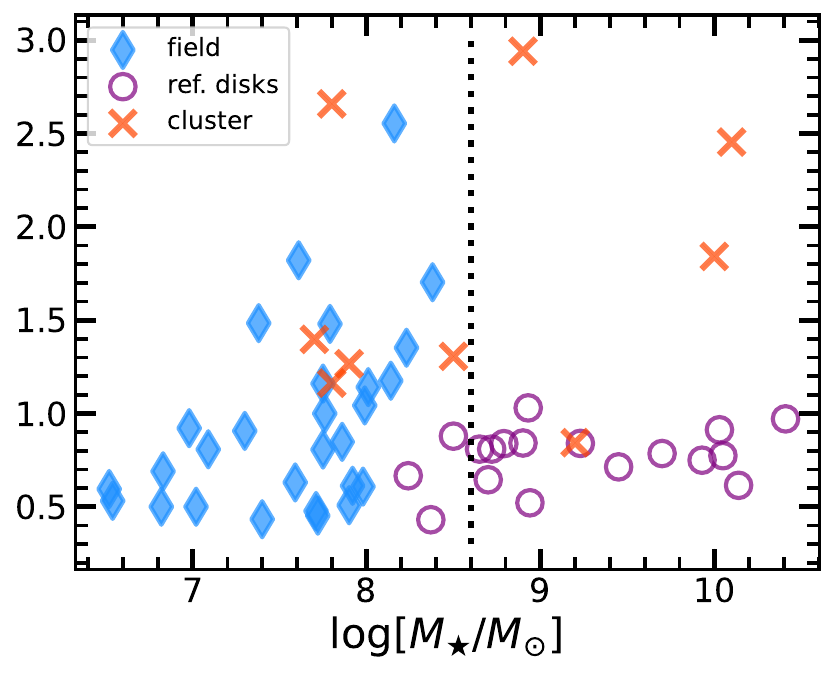} 
    \caption{$D_{\rm stellar}/D_{\HI{}}$ as a function of $M_{\HI{}}$ (left) and $M_{\star}$ (right) for galaxies in the sample where both size measurements are available. The vertical dotted line in the right panel marks the location of $m_{break}$ in the stellar size--mass relation.}
    \label{fig:Dstellar_DHI_ratio}
\end{figure*}

\begin{figure}[h]
    \centering
    \includegraphics[width=0.5\textwidth]{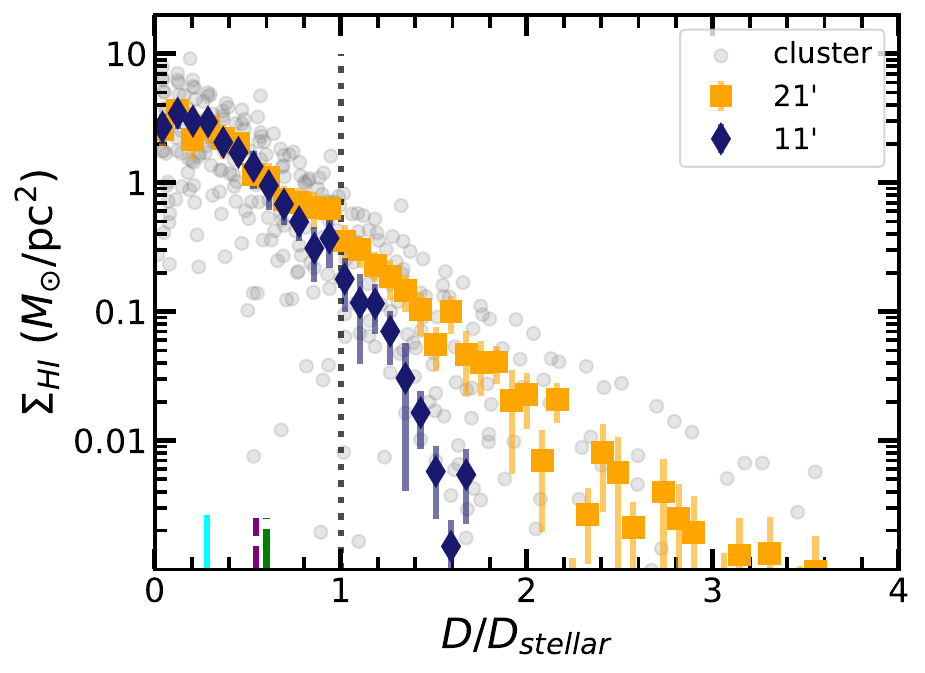}
    \caption{MeerKAT \HI{} surface density profiles of Fornax cluster galaxies, normalized using $D_{\rm stellar}$. As before, the symbol `D' refers to diameter ($2\times$ radius). The mean profile using the 11\arcsec (blue diamonds) and 21\arcsec (orange squares) resolution \HI{} maps obtained by averaging the profiles from individual galaxies (grey dots) are plotted for comparison. The largest value of ratio between the beam size $\sqrt{B_{maj} \times B_{min}}$ and $D_{\rm stellar}$ in this sample are represented as the cyan and purple ticks in the x-axis, confirming that we resolve \HI{} at $D_{\rm stellar}$. $D_{\HI{}}$ appears at 0.6\,$D_{\rm stellar}$ in this axis (green tick). The \HI{} surface density near $D_{\rm stellar}$ is $\sim 0.3\,M_{\odot}/$pc$^2$, more than a factor of three lower compared to the $1\,M_{\odot}$/pc$^2$ contour. These profiles quantify the illustration previously shown in Fig. \ref{fig:mfs}.}
    \label{fig:hi_profiles}
\end{figure}


\item Fifth, the $D_{\rm stellar}$ of \HI{} warped edge-on galaxies from the literature in the higher mass regime are scattered around their $D_{\HI{}}$ values. These values are also compatible with the location where the warp features start \citep[i.e. $\sim$ 1.1\,$D_{\rm stellar}$; see][]{vanderkruit2007aap466_883}. However, the extreme low surface brightness galaxies \emph{UGC~1382} and \emph{Nube} are above the $D_{\HI{}}-M_{\HI{}}$ relation. 
\end{itemize}

\section{Discussion} \label{sect:discussion}



We have shown that $D_{\rm stellar}$ is not always within $D_{\HI{}}$. This result is particularly visible for the Fornax cluster galaxies in our sample, even though \HI{} has been detected well beyond the stellar edge (Fig. \ref{fig:mfs} and \ref{fig:hi_profiles}). The fact that low density \HI{} is observed beyond $D_{\rm stellar}$ poses a problem in how \HI{} sizes are currently represented in scaling relations. As $D_{\HI{}}$ is defined at the $1\,M_{\odot}$/pc$^2$ level, the isophote is not representative of lower density material and thus cannot be taken at face value in any comparison with stellar sizes. If $D_{\HI{}}$ was measured at an even lower density of, for example, $0.5\,M_{\odot}/$pc$^2$, $\sim 10$\% more \HI{} mass would be enclosed and such a measure would result in a $D_{\HI{}} - M_{\HI{}}$ relation with the same slope but a larger scatter \citep[see Fig. 4 in][]{wang+2016mnras460_2143}. The similar slope can be explained by the homogeneity of \HI{} profiles in the outskirts \citep[e.g.][]{wang+2014mnras441_2159}. Therefore, simply using a lower density threshold that is within the resolution and sensitivity limits of current \HI{} surveys may not necessarily solve the problem of how to best encapsulate \HI{} size. 


Although $D_{\HI{}}$ has units of length, it is more informative to interpret the scaling relation not in terms of `\HI{} size' but rather on what $D_{\HI{}}$ actually traces, i.e. the location of the surface density of gas in which $\Sigma_{\HI{}} = 1\,M_{\odot}$/pc$^2$. In other words, $D_{\HI{}}$ is a reference for how densely distributed or concentrated the \HI{} gas is found in a galaxy which directly impacts star formation \citep{wang+2020apj890_63, yu+2022apj930_85} . Within this context, when $D_{\rm stellar} >> D_{\HI{}}$, it is either because the galaxy is isolated with extremely diffuse \HI{} at its current stage of evolution \citep[e.g.][]{hogg+2007aj134_1046, leisman+2017apj842_133} or else because an environmental process has lowered the density of gas across the stellar region, making the \HI{} more concentrated \citep[e.g.][]{lee+2022mnras517_2912, loni+2023mnras523_1140}. 
We discuss the two key results of this work in the sections below, additionally using the observation that galaxies are generally found within $3\sigma$ of the $D_{\HI{}}-M_{\HI{}}$ relation despite their stage of evolution or the presence of external forces which can change their \HI{} content \citep{stevens+2019mnras490_96}. \par 


\subsection{The break in the $D_{\rm stellar}-M_{\star}$ relation} 
\label{sect:break}
 
If the edge of a galaxy is formed in situ due to a star formation threshold \citep{trujillo+2020mnras493_87}, it has been found that this threshold varies across stellar mass and morphology \citep{chamba+2022aap667_87} and with environment \citep{chamba+2023arXiv2311.10144}. 
For spirals and dwarf galaxies, it is interesting to speculate that the broken relation found in the $D_{\rm stellar}-M_{\star}$ plane at $M_{\star} \sim 4\times10^{8}$ $M_{\odot}$ potentially distinguishes the dominant physical processes regulating size between these two families of galaxies. \par 

The idea that the impact of different physical or evolutionary processes are responsible for a change in slope in scaling relations is not new. For example, \citet{derijcke+2009mnras393_798} argued that the insensitivity of \citet{sersic1968book}-based photometric scaling relations (i.e. S\'ersic indices, effective radii, see Introduction) to the environment of a sample of dwarf and massive early-type galaxies, combined with the observation of a change in slope in their scaling relations (e.g. effective radius vs. absolute magnitude, see their Fig. 2) imply evolutionary processes such as supernova (or stellar) feedback which dominate at different mass regimes (discussed later in Sect. \ref{sec:feedback}).\par
The more recent study by \citet{watkins+2023mnras521_2012} on the $3.63\,M_{\odot}$/pc$^2$ isophotal radius--stellar mass relation also suggests the existence of a dichotomy in the structures of dwarfs and massive galaxies. However, as previously mentioned in the Introduction, the high scatter in the relation at lower stellar masses complicates interpretation as this radius measure cannot be applied to all diffuse galaxies, similar to the issue with $D_{25}$ shown in Fig. \ref{fig:D25_examples}. This issue is briefly demonstrated in Appendix \ref{app:isophote} where the size--mass relation using stellar edges and the \citet{watkins+2023mnras521_2012} isophote are compared.\par 

Moreover, from a cosmological perspective, how the relationship between galaxy stellar and halo mass changes and drives evolutionary processes in the dwarf regime is also actively investigated in the literature \citep[e.g.][]{sawala+2010mnras402_1599, hopkins+2014mnras445_581, garrisonkimmel+2017mnras464_3108, wechsler+2018araa56_435, rey+2022mnras511_5672, azartashnamin+2024apj970_40}. Therefore, it is interesting to consider the properties of stellar edges as a mass dependent, observable diagnostic of structure assembly via processes such as stellar accretion (Sect. \ref{sec:accretion}), feedback (Sect. \ref{sec:feedback}), migration (Sect. \ref{sec:migration}) and environmental influences (Sect. \ref{sect:environment}). The discussion of these different physical processes and their potential association with $m_{break}$ follows.

\subsubsection{Inside-out growth via accretion}
\label{sec:accretion}

Above $m_{break}$, the $D_{\rm stellar}-M_{\star}$ relation is populated by more massive disk galaxies that can accrete stars and gas from the surrounding interstellar medium as well as neighbouring galaxies i.e. the inside-out formation scenario.  \citet{chamba+2022aap667_87} reports that the truncations of disk galaxies with $M_{\star} > 10^9\,M_{\odot}$ scale with a slope closer to $\sim M_{\star}^{1/3}$. This slope is also reflected in the data points of disk galaxies taken from the literature (Fig. \ref{fig:dedge_mstar}). In particular, the warped \HI{} distribution of the \citet{vanderkruit2007aap466_883} edge-on galaxies do not make them outliers (i.e. the data points are within 3$\sigma_{obs}$ of the observed relation), even though the edge locations may shift by 1-2\,kpc due to its presence \citep[see][]{martinezlombilla+2023aap678_62}. These results show that the scaling relation is robust even if truncations can be affected by both a star formation threshold and the presence of the warp. \par

In the case of galaxies UGC~00180 \citep{trujillo+2016apj823_123} and NGC~1042 \citep{trujillo+2021aap654_40} which have stellar halo light, the truncation is also found near the threshold for star formation $\Sigma_{\star} \sim 1-2\,M_{\odot}$/pc$^2$ \citep{schaye2004apj609_667, trujillo+2020mnras493_87}. Subsequently, the ex situ stellar halo component is formed via the accretion of satellite galaxies \citep[e.g.][]{cooper+2010mnras406_744}. The fact that these galaxies are also not outliers but rather follow the size-mass scaling relation suggests that the truncation likely formed first as part of the disk, setting a marker for the onset of the ex situ stellar halo. \par

\subsubsection{Stellar Feedback}
\label{sec:feedback}

Interestingly, $m_{break}$ is within the $M_{\star}$ range where stellar feedback is predicted to be most influential in isolated galaxies \citep[i.e. $\sim 10^7-4\times10^{9}\,M_{\odot}$;]{elbadry+2016apj820_131, emami+2021apj922_217}. Stellar feedback causes outflows of gas which can then cool and fall back into the center of the galaxy, inducing star formation. In the process, these gaseous outflows may also lower the gravitational potential significantly, promoting the radial migration of stars outwards. Therefore, the feedback process could change both the \HI{} and stellar sizes\footnote{In this context, the phrase `stellar size' is used to refer to traditional size measures of the stellar light in galaxies and not specifically to $D_{\rm stellar}$.}. \par 


Below $m_{break}$, the above effects potentially explain the small scatter and higher slope of the $D_{\rm stellar}-M_{\star}$ relation. \citet{elbadry+2016apj820_131} show that while the effective radius fluctuates significantly by factors of 2--3 over 100\,Myr timescales due to feedback, the radius enclosing 90\% of $M_{\star}$ ($R_{90}$) barely changes (see their Fig. 2 and 8). $R_{90}$ is used by these authors as a proxy for the location of the global stellar distribution. Therefore, it is reasonable to expect that $D_{\rm stellar}$ would also not significantly change due to feedback and the scatter of the $D_{\rm stellar}-M_{\star}$ relation is not significantly affected. \par

If we exclude the extreme low surface brightness galaxies, all but one of our \HI{} sample are of late-type morphology. The higher slope of $\sim 0.45$ in the low mass regime where $M_{\star} < m_{break}$ is found when also including early-type galaxies without gas in the \citep{chamba+2023arXiv2311.10144} sample, indicating that morphology or gas-richness are not the main drivers of the higher slope.  Rather, the higher slope below $m_{break}$ is more likely a combined reflection of the rapid evolution and changes occurring in the stellar and \HI{} distribution, masses and total gravitational potential of dwarf galaxies due to feedback as well as environmental processes, reflecting the possibility of both inside-out and outside-in formation scenarios in dwarfs \citep[see][]{chamba+2022aap667_87, rey+2022mnras511_5672, zhu+2024arXiv2404.00129, azartashnamin+2024apj970_40}. The impact of environmental processes is discussed in more detail in Sect. \ref{sect:environment}. However, we point out that the relation is incomplete towards lower mass, highly extended low surface brightness galaxies such as \emph{Nube}. Such galaxies could potentially lower the slope we report here. \par

The above interpretation for existence of the break at $m_{break}$ and thus the higher slope is also supported by structure of the $D_{\rm stellar}-M_{\HI{}}$ plane. 
The $M_{\HI{}} < 10^9\,M_{\odot}$ sample displays $\sim 39\%$ more scatter in the $D_{\rm stellar}-M_{\HI{}}$ plane relative to higher mass galaxies, demonstrating 
that stellar edges of low mass field galaxies can be found either within or beyond $D_{\HI}$ at a fixed $M_{\HI{}}$ (Fig. \ref{fig:Dedge_MHI}).  
Assuming galaxies redistribute and consume $\HI{}$ gas as a consequence of feedback-induced star formation such that they are still along the $D_{\HI}-M_{\HI{}}$ relation \citep[see][]{stevens+2019mnras490_96}, the large variation of $D_{\rm stellar}$ with respect to $D_{\HI{}}$ at a fixed $M_{\HI{}}$ is likely an imprint of this active star formation process. Consequently, the slope of the $D_{\rm stellar}-M_{\HI{}}$ plane is not parallel to the  $D_{\HI{}}-M_{\HI{}}$ relation as was the case for $D_{25}-M_{\HI{}}$ \citep{broeils+1997aap324_877}.\par

\subsubsection{Stellar Migration}
\label{sec:migration}

In the discussion on stellar feedback above, it was briefly mentioned that feedback-induced stellar migration can occur in isolated galaxies. However, based on the results shown in \citet{elbadry+2016apj820_131}, it is argued that $D_{\rm stellar}$ may not significantly change due to ongoing feedback even as $D_{\HI{}}$ (and $M_{\HI{}}$) changes and this scenario provides an explanation for the distribution of galaxies in the $D_{\rm stellar}-M_{\HI{}}$ plane (Sect. \ref{sec:feedback}). As stellar migration is a physical process in galaxies, it is interesting to consider if it is specifically associated with $m_{break}$. There are three observational considerations which support the idea that stellar migration may not be the dominant effect causing the break in the size--mass relation. \par 

First is that the diversity of the shapes of color profiles in dwarf galaxies \citep[e.g.][]{herrmann+2016aj151_145, chamba+2022aap667_87} where the outer regions can either be bluer or redder due to their diverse star formation histories \citep[i.e. inside-out and outside-in formation scenarios, see e.g.][]{koleva+2009aap501_1269, zhang+2012aj143_47} as well as the metallicity gradients observed in dwarfs \citep[e.g.][]{schroyen+2013mnras434_888, taibi+2022aap665_92} supports the idea that stellar migration (induced by e.g. stellar feedback) occurs widely in these galaxies. But stellar migration can also occur in massive late-type disk galaxies and can explain their typical ``U'' shaped color profile  \citep[][and references therein]{bakos+2008apj683_103, marino+2016aap585_47, debattista+2017inproceedings_77}. Although the processes causing the migration differ, e.g. due to bars and spiral arm resonances in massive disk galaxies, the observation of stellar migration in both dwarfs and spirals shows that the process is not particularly unique to either family of galaxies and occurs widely. \par 

Second is the observation of a color bifurcation or stratification in the size--mass plane by \citet{chamba+2022aap667_87, chamba+2023arXiv2311.10144} where at a fixed stellar mass, galaxies with bluer edges (and blue average color) have larger sizes compared to redder ones. This bifurcation is found along the full range of galaxies in stellar mass  $10^5\,M_{\odot} <M_{\star} < 10^{11} \,M_{\odot}$. If stellar migration was the dominant process regulating the sizes of massive disk galaxies for example, then the opposite trend is expected i.e. galaxies with redder edges would be larger as the older stars in the inner regions of galaxies migrate outwards. But this scenario is not observed.\par 

Moreover, bluer and younger stars move $\sim$1\,kpc at most due to migration \citep[see][]{elbadry+2016apj820_131} which is not large enough to explain the color gradient in the size--mass plane. Therefore, the dominant effect driving the bifurcation is more likely a combination of (1) the environment, where in groups and clusters outside-in quenching occurs due to the removal of gas in the outer regions of galaxies, thus lowering the \HI{} density within $D_{\rm stellar}
$ (Fig. \ref{fig:hi_profiles} and further discussed in Sect. \ref{sect:environment}) and consequently (2) the morphology of galaxies where early-type galaxies have been found to be smaller than late-type ones regardless of their environment \citep[see Fig. 14 and 15 in][]{chamba+2023arXiv2311.10144} and (3) stellar feedback which regulates the surface density of \HI{} in galaxies (Sect. \ref{sec:feedback}), consistent with the \citet{chamba+2023arXiv2311.10144} observation that galaxies with more dense stellar edges (which are also redder, see their Fig. 8) are also those with lower \HI{} fractions (their Fig. 16, mentioned earlier in the Introduction).

Third is a potential bias in the bifurcation result and its applicability because the sample used in this work or in \citet{chamba+2023arXiv2311.10144} does not consist of gas-rich, extended, red, isolated low surface brightness dwarf galaxies. The distances of these galaxies are difficult to determine in large numbers, but surveys such as from Euclid and the future Roman and SKA will be able to probe such galaxies. Future work will be able to investigate if these redder low surface brightness galaxies have varied \HI{} sizes similar to the star forming isolated galaxies. The result from such an analysis could provide more clear evidence on the role of stellar migration in shaping the sizes of galaxies.




\subsection{Outside-in quenching via environmental processes in groups and clusters}
\label{sect:environment}

As discussed in \citet{chamba+2023arXiv2311.10144}, 
tidal forces from the gravitational potential of the Fornax Cluster play a  more prominent role in shaping the properties of dwarf member galaxies  \citep[e.g.][]{asencio+2022mnras515_2981} than do those involved in galaxy-galaxy interactions. 
Together, gas removal in the cluster environment due to ram pressure and the cluster potential can lower the \HI{} density within the stellar boundary of galaxies \citep[i.e. outside-in quenching, see also e.g.][]{morokumamatsui+2022apj263_40,watts+2023pasa40_017, loni+2023mnras523_1140}. Consequently, this process could make $D_{\HI} < D_{\rm stellar}$, such as is observed in this work. \par 
In the MeerKAT Fornax cluster sample analyzed here, the \HI{} surface density near $D_{\rm stellar}$ is typically $\sim 0.3\,M_{\odot}/$pc$^2$ which is more than a factor of three lower compared to the $1\,M_{\odot}$/pc$^2$ level. This value is even lower compared to the expected \HI{} star formation threshold for dwarfs \citep[e.g.][]{begum+2008mnras386_1667} and is further evidence of the lowering of \HI{} densities within the stellar boundaries of galaxies due to the cluster environment. Given that we currently do not have deep \HI{} maps for all the field galaxies in the \citet{chamba+2023arXiv2311.10144} sample (Table \ref{tab:data_source}), a more detailed analysis on how $D_{\rm stellar}$ is related to the star formation threshold in \HI{} surface density will be performed in future work (Chamba et al. in prep). \par 

Consequently, all except one of the Fornax cluster galaxies follow the $D_{\HI{}}-M_{\HI{}}$ relation and are systematically located in the upper half of the $D_{\rm stellar}-M_{\HI{}}$ plane. In other words, their $D_{\rm stellar} >> D_{\HI{}}$ at a fixed $M_{\HI{}}$. The physical interpretation of this result is explained in the following. If the location of $D_{\rm stellar}$ is fixed soon after the dwarf galaxy has been rapidly quenched via gas removal due to cluster tides and ram pressure \citep{kleiner+2023aap675_108, chamba+2023arXiv2311.10144}, then it is reasonable to assume that any future loss of gas via processes that specifically act only on gas (e.g. ram pressure) as described in e.g. \citet[][]{stevens+2019mnras490_96} will not further change the location of $D_{\rm stellar}$. 



As discussed in \citet{chamba+2023arXiv2311.10144}, such future loss of gas is a possibility because the time scale of gas removal via a process like ram pressure is $\sim 1-2$\, Gyr \citep[see][]{loni+2021aap648_31} which is much smaller than the typical $\sim 10$\, Gyr age of a dwarf in Fornax \citep{rakos+2001aj121_1974, jerjen2003aap398_63}. But $D_{\rm stellar}$ will not undergo further change because (1) there are no hints of tidal features on the stellar bodies of the galaxies studied in \citet{chamba+2023arXiv2311.10144} and (2) as explained earlier, the cluster tides and ram pressure first impacts the outer low density \HI{} gas component so $D_{\rm stellar}$, once formed, will not be impacted further by these processes that impact gas. Along with these considerations, the finding of this work that $D_{\rm stellar} >> D_{\HI{}}$ at a fixed $M_{\HI{}}$ suggests that the gas detected Fornax members are still in the active process of converging to their final \HI{} sizes due to the additional, environmental processes acting on them after in-fall \citep[see also][]{derijcke+2010apj724_171} .


This interpretation is supported when also considering the distribution of field galaxies in the scaling relations. While not as extreme as in a cluster environment, ram pressure in group environments can also quench surrounding satellite galaxies \citep[e.g.][]{penny+2016mnras462_3955, hausammann+2019aap624_11, zhu+2023mnras521_3765}. The fact that we find group satellites both above and below the $D_{\rm stellar
}-M_{\HI{}}$ relation is likely an imprint of the intermediate ram pressure felt by the satellite from the host halo, similar to the results discussed in \citet{chamba+2023arXiv2311.10144}. On the other hand, the similar distribution of isolated galaxies in the $D_{\rm stellar
}-M_{\HI{}}$ relation compared to the satellites is potentially a consequence of feedback (see Sect. \ref{sect:break}).   \par 

We remind the reader that the early-type dwarfs in Fornax which are devoid of gas do not appear in the scaling relations discussed in this work. The stellar edges of those galaxies are the smallest in comparison to the gas-rich late-type dwarfs shown here, likely a reflection of the impact of gas removal outside-in \citep[e.g.][]{bekki2009mnras399_2221} which can rapidly quench galaxies. In any case, all these findings are consistent with the idea that if $D_{\rm stellar}$ does not undergo further change as explained earlier, then continuous gas removal in the cluster could make $D_{\HI{}} << D_{\rm stellar}$ until all the gas runs out \citep[see also][]{stevens+2019mnras490_96}. In other words, the movement of \HI{} in galaxies along the $D_{\HI{}} - M_{\HI{}}$ is consistent with their distribution in the $D_{\rm stellar} - M_{\HI{}}$ plane. \par 
However, more diffuse galaxies like \emph{Nube}, UGC~1382 or galaxies with higher $\HI{}$ masses where $D_{\HI{}} >> D_{\rm stellar}$ could challenge these results as they could tilt the slope of the $D_{\rm stellar} - M_{\HI{}}$ relation towards higher values. Investigating this issue in future work will be readily facilitated by combining the deeper observations expected from the Nancy Grace Roman Space Telescope \citep[e.g.][]{montes+2023arXiv2306.09414} and the Square Kilometer Array \citep[e.g.][]{dutta+2022afz43_103}. \par

\section{Conclusions}

By comparing the location of the stellar edges of galaxies $D_{\rm stellar}$ where in situ star formation cuts off with the \HI{} diameter defined at the 1\,$M_{\odot}$/pc$^2$ \HI{} surface density, $D_{\HI}$, we have found that $D_{\rm stellar}$ is not always less than $D_{\HI{}}$ as is commonly assumed for gas-rich disk galaxies. When considering field environments, the larger variation in the $D_{\HI{}}/D_{\rm stellar}$ for low mass $M_{\HI{}} < 10^9\,M_{\odot}$ dwarfs compared to more massive disks reflects the influence of stellar feedback which can rapidly change the concentration of \HI{} and therefore the location of $D_{\HI{}}$. For the Fornax cluster dwarf sample, the observation that their $D_{\rm stellar} >> D_{\HI{}}$ , combined with the much lower \HI{} surface densities found within the $D_{\rm stellar}$ of these galaxies, shows the impact of outside-in gas removal and quenching due to ram pressure and cluster tides in their higher density environment. \par 
The break in the size--stellar mass relation characterized with a higher slope for galaxies where $M_{\star} < 4\times 10^8\,M_{\odot}$ suggests that lower mass galaxies regulate their sizes more rapidly due to feedback and environmental processes. On the other hand, stellar and \HI{} accretion in more massive disk galaxies do not significantly alter the scatter in the size--stellar mass relation and their $D_{\rm stellar} \lesssim D_{\HI{}}$ as is commonly assumed in the literature. Combining future deep imaging surveys in the optical and \HI{} from the LSST-Rubin, Euclid, Roman and the SKA will facilitate the applicability of the methods explored here to larger samples of galaxies, including the low surface brightness and \HI{} poor.

\begin{acknowledgments}

We thank the anonymous referee for their comments and suggestions which improved the clarity of our manuscript. N. C., V. J. M. L. and S. D. C.'s research are supported by an appointment to the NASA Postdoctoral Program at the NASA Ames Research Center, administered by Oak Ridge Associated Universities under contract with NASA. N. C. additionally acknowledges support for the collaborative visits funded by the Cosmology and Astroparticle Student and Postdoc Exchange Network (CASPEN). 
M. J. H.  is Fellow of the Knut \& Alice Wallenberg Foundation and acknowledges support from the research project grant ‘Understanding the Dynamic Universe’ funded by the Knut and Alice Wallenberg Foundation under Dnr KAW 2018.0067.


\end{acknowledgments}

%



\newpage
\appendix

\section{Fitting $D_{\rm stellar}-M_{\star}$}

\begin{figure}[h]
    \centering
    \includegraphics[width=0.5\linewidth]{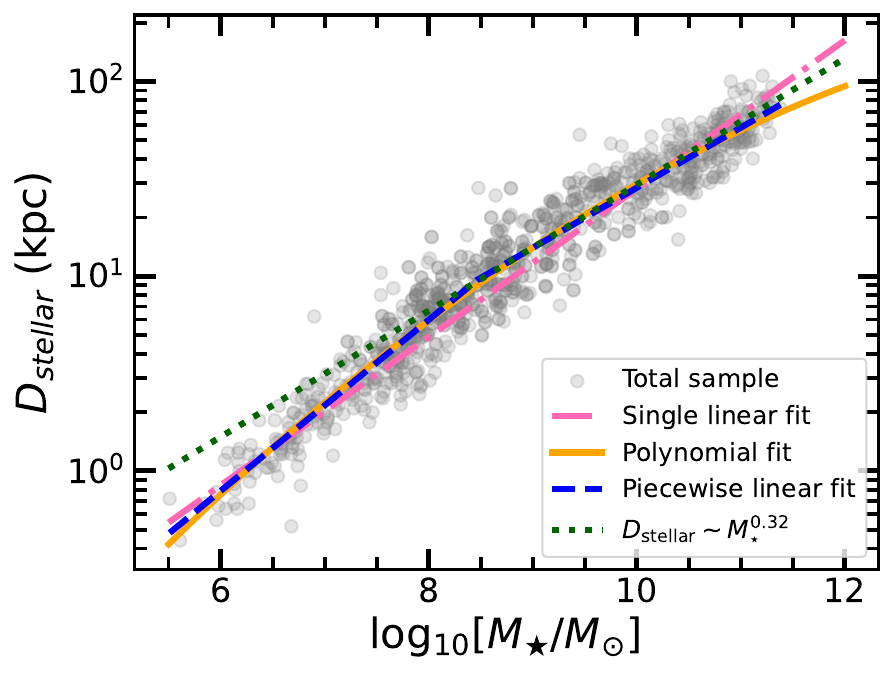}
    \caption{Fitting $D_{\rm stellar}-M_{\star}$ using different functional forms. The deviation of the size--mass relation from the $D_{\rm stellar}\sim M_{\star}^{0.32}$ line below $m_{break} \sim 4 \times 10^{8}\,M_{\odot}$ is visible when considering the piecewise linear and polynomial fit. It is this deviation we refer to as a break in the size--mass relation.}
    \label{fig:functional_form}
\end{figure}

To further demonstrate the significance of the location of the break at $m_{break}$ in the size--mass relation, Fig. \ref{fig:functional_form} compares the best fit piecewise linear relation with a single linear and second order polynomial function. The $D_{\rm stellar}\sim M_{\star}^{0.32}$ relation from \citet{chamba+2022aap667_87} is also shown to highlight the location of $m_{break} \sim 4 \times 10^{8}\,M_{\odot}$ where galaxies with lower masses follow a size--mass scaling relation with higher slope. We chose the piecewise linear relation given the very low non-linear p-value associated with $m_{break}$ and for comparison with the previous work on galaxy edges. However, even if the relation can be represented with a continuous second order polynomial, the deviation of the size--mass relation from the $D_{\rm stellar}\sim M_{\star}^{0.32}$ line below $m_{break}$ is visible. It is this deviation we refer to as a break in the size--mass relation.  

\section{Complications with using isophotal radii in scaling relations}
\label{app:isophote}

\begin{figure}
    \centering
    \includegraphics[width=0.5\linewidth]{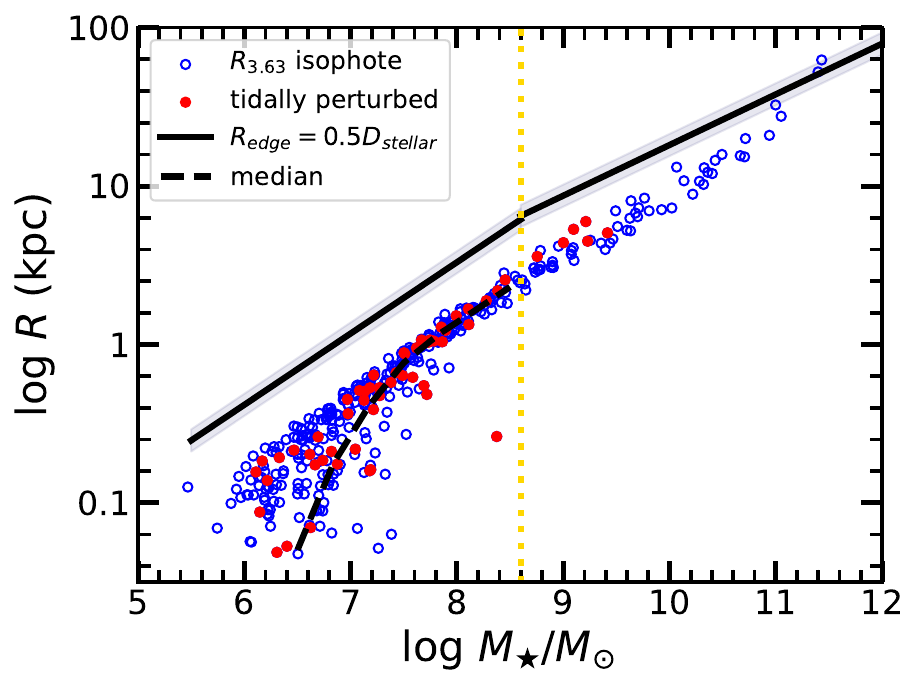}
    \caption{Radius measured at a fixed isophote at 3.63\,$M_{\odot}$/pc$^2$ by \citet{watkins+2023mnras521_2012}, called $R_{3.63}$ (blue circles) for Fornax cluster dwarf galaxies and a sample of nearby disk galaxies. The Fornax galaxies with visually identified tidal perturbations in their stellar bodies from \citet{venhola+2022aap662_43} are plotted in red. The sigma-clipped median (dashed line) is shown for galaxies with $M_{\star} < 10^{9}\,M_{\odot}$.  For comparison, we show the broken radius size--stellar mass relation as defined using stellar edges ($R_{\rm edge}$) and the shaded region shows the intrinsic scatter of 0.07\,dex for reference. The yellow vertical dotted line is the location of $m_{break}$. This plot is yet another example of the complications in interpreting isophotal scaling relations as `size' relations because the radius does not represent the same regions of different galaxies which increases the scatter and in this case, creates a curved structure in the relation. We refer the reader to \citet{trujillo+2020mnras493_87} and \citet{chamba+2020aap633_3} where the impact of using isophotal radii as size definitions has also been discussed at length.}
    \label{fig:r3.63}
\end{figure}

As mentioned in the Introduction, \citet{watkins+2023mnras521_2012} use a fixed isophotal radius measure at the stellar surface density of 3.63\,$M_{\odot}$/pc$^2$ (denoted as $R_{3.63}$). These authors show that the relation curves as lower mass dwarf galaxies tend towards zero sizes. In this appendix, we demonstrate this issue in Fig. \ref{fig:r3.63} by plotting the $R_{3.63}$ vs. $M_{\star}$ plane using the data points catalogued in Watkins et. al. and comparing the resulting scaling relation with that using stellar edges from this work. Several remarks are in order.

First, the $R_{3.63}$ relation is systematically below the stellar edge one. This result is simply explained by the fact that the 3.63\,$M_{\odot}$/pc$^2$ contour is at a higher surface density than the stellar mass surface density at the location of the edges \citep[$\sim $0.5--3\,$M_{\odot}$/pc$^2$;see][]{chamba+2022aap667_87}. As the radial surface density profiles of galaxies decrease over increasing radius, it is reasonable to assume that higher surface densities occur in the inner regions of galaxies. Therefore, smaller $R_{3.63}$ at a fixed $M_{\star}$ will tend towards the central regions of dwarfs, and zero if the surface density is never reached.

Second, Watkins et al. explore even lower surface densities as their radius in their work, and as expected, the curvature of the relation in the dwarf regime disappears. This result is further indication of the sensitivity of isomass radii to structural changes in galaxies over stellar mass and is one of the main arguments in their work supporting the existence of a \textit{structural} dichotomy in the relation which separates dwarfs and massive galaxies. While this conclusion is also supported by the work shown here, we emphasise that the results from Watkins et al. (or from the earlier work by \citet{derijcke+2009mnras393_798} using the effective radius) cannot be used as evidence for a break in the size--stellar mass relation as reported in this paper. This statement is supported by the simple fact that $R_{3.63}$ may not be representative of the stellar boundaries of dwarf galaxies due to the effect described in the first point above and the Introduction on the limitations of traditional effective and isophotal size definitions. \par 

Third, it is difficult to physically interpret the curvature of the relation. While the curvature is shown to be a direct consequence of the \citet{sersic1968book} model which is used to describe the light distribution in galaxies, and is thus also consistent with the \citet[][]{sanchezalmeida2020mnras495_78} analysis, by ``physical'' in this context, we are referring to the interpretation of scaling relations in the context of galaxy evolution theory. We demonstrate this issue by considering the proposal in \citet{watkins+2023mnras521_2012} who argue that the outliers of the relation, defined as those approximately below the median relation shown here (but see the exact selection criteria in their Fig. 7), are likely those that ``formed through tidal disturbances'' in the Fornax Cluster. However, as already discussed in \citet{chamba+2023arXiv2311.10144}, they do not provide any further evidence or metrics to verify if these outliers are tidally perturbed which is an observable signature of past tidal interactions or formations.\par 

Tidally perturbed galaxies often show features such as tails or asymmetries in or outside their stellar bodies. If tidally perturbed galaxies are outliers in this scaling relation, the expectation would be to find tidal features in a large fraction of these outliers. In Fornax, \citet{venhola+2022aap662_43} visually searched for such features in the cluster for each galaxy and classified them as ``regular'', ``slightly disturbed'' or ``clearly disturbed''. Fig. \ref{fig:r3.63} additionally labels the slightly and clearly tidally perturbed galaxies collectively as red points in the \citet{watkins+2023mnras521_2012} sample. The figure shows that (1) most of the visually disturbed galaxies appear above the median relation in the low mass regime $M_{\star} < 10^{7}\,M_{\odot}$ and (2) only a small fraction of the outliers appear perturbed in the criteria of \citet{venhola+2022aap662_43}. \par 

This result shows that not all outliers in the $R_{3.63}$ relation visibly appear to have tidal features. While it is possible that not all tidal features were detected by \citet{venhola+2022aap662_43}, this result demonstrates that it is more likely that the dependence of $R_{3.63}$ on the varied light distribution of galaxies causes the curvature in the scaling relation and not tidal disturbances. While this appendix specifically discusses the recent work by \citet{watkins+2023mnras521_2012} to demonstrate the complications with using and interpreting isophotal radii in scaling relations, no matter which radius measure is used by authors, this analysis highlights the importance of accounting for potential biases of the chosen parameters when addressing the origin of scaling relations.




\bibliography{pruned.bib}
\bibliographystyle{aasjournal}




\end{document}